\documentclass[aps,prc,amsfonts,preprintnumbers,superscriptaddress,showpacs,nofootinbib]{revtex4}

\usepackage{graphics}
\usepackage{psfrag}
\usepackage{epsfig}
\usepackage{epsf}
\usepackage{float}

\newcommand{\fet}[1]{\mbox{\boldmath $#1$}}

\newcommand{\beq}{\begin{equation}}
\newcommand{\eeq}{\end{equation}}
\newcommand{\beqa}{\begin{eqnarray}}
\newcommand{\eeqa}{\end{eqnarray}}
\newcommand{\nn}{\nonumber \\ }

\begin{document}

\preprint{{\tiny FZJ-IKP-TH-2007-33}}
\preprint{{\tiny HISKP-TH-07/27}}

\title{Subleading contributions to the chiral three-nucleon 
force I:\\[0.2em]
long-range terms}

\author{V.~Bernard}
\email[]{Email: bernard@lpt6.u-strasbg.fr}
\affiliation{Universit\'e Louis Pasteur, Laboratoire de Physique Th\'eorique
  3-5, rue de l'Universit\'e, F-67084 Strasbourg, France}
\author{E.~Epelbaum}
\email[]{Email: e.epelbaum@fz-juelich.de}
\affiliation{Forschungszentrum J\"ulich, Institut f\"ur Kernphysik 
(Theorie), D-52425 J\"ulich, Germany}
\affiliation{Universit\"at Bonn, Helmholtz-Institut f{\"u}r
  Strahlen- und Kernphysik (Theorie), D-53115 Bonn, Germany}
\author{H.~Krebs}
\email[]{Email: hkrebs@itkp.uni-bonn.de}
\affiliation{Universit\"at Bonn, Helmholtz-Institut f{\"u}r
  Strahlen- und Kernphysik (Theorie), D-53115 Bonn, Germany}
\author{Ulf-G.~Mei{\ss}ner}
\email[]{Email: meissner@itkp.uni-bonn.de}
\homepage[]{URL: www.itkp.uni-bonn.de/~meissner/}
\affiliation{Universit\"at Bonn, Helmholtz-Institut f{\"u}r
  Strahlen- und Kernphysik (Theorie), D-53115 Bonn, Germany}
\affiliation{Forschungszentrum J\"ulich, Institut f\"ur Kernphysik 
(Theorie), D-52425 J\"ulich, Germany}
\date{\today}

\begin{abstract}
We derive the long-range contributions to the tree-nucleon force at
next-to-next-to-next-to-leading order in the chiral expansion. We give
both momentum and coordinate space representations.
\end{abstract}


\maketitle

\vspace{-0.2cm}

\section{Introduction}
\def\theequation{\arabic{section}.\arabic{equation}}
\label{sec:intro}

Three-nucleon forces (3NFs) are an  indispensable ingredient in accurate
few-nucleon and nuclear structure calculations. In particular, the
three-nucleon system shows clear evidence for 3NFs, see the recent general
introduction \cite{KalantarNayestanaki:2007zi}.
Chiral effective field theory is the appropriate tool to analyze nuclear forces.
Precise two-nucleon potentials have been developed at
next-to-next-to-next-to-leading order (N$^3$LO) in the chiral expansion, see
Refs.~\cite{Entem:2003ft,Epelbaum:2004fk}. 3NFs first appear at N$^2$LO in
the Weinberg counting scheme  \cite{Weinberg:1990rz,Weinberg:1991um}
and have been analyzed and scrutinized in
\cite{vanKolck:1994yi,Friar:1998zt,Epelbaum:2002vt,Epelbaum:2005pn,Nogga:2005hp,Navratil:2007we}.
There are various
reasons to improve the theoretical precision of these 3NFs: 1) one should
utilize the two- and three-nucleon forces at the same order in the expansion,
2) there are some outstanding discrepancies between theory and experiment
like e.g the recently measured differential cross section in deuteron break-up 
at low energies~\cite{Ley:2006hu} or the long-standing $A_y$ puzzle \cite{Koike,Witala}
 and 3) the theoretical uncertainty employing only the
leading 3NFs quickly grows with increasing energy if one investigates e.g.
nucleon-deuteron scattering or break-up. It is therefore timely and necessary
to derive the chiral 3NF at N$^3$LO. In this work, we focus our attention
on the long-range contributions at this order, which are free of unknown 
coupling constants. Here, we consider the chiral effective Lagrangian with 
pions and nucleons. The precise relation of the
results presented here to an effective field theory including also spin-3/2
degrees of freedom will be the subject of a subsequent paper.

As we will show later, there are five different topologies contributing to the
3NF at N$^3$LO. From these, three topologies make up the long-range contribution,
which is defined by not including any multi-nucleon contact interactions. 
This long-range part is given by two-pion exchange ($2\pi$) graphs, 
two-pion--one-pion exchange ($2\pi$-$1\pi$) graphs and the so-called ring diagrams,
where the pion loop connects all three nucleon lines. The first type of graphs
has recently been considered based on the so-called infrared regularization
in Ref.~\cite{Ishikawa:2007zz}, we will compare our results to that work
below. The spin-isospin structures originating from ring diagrams with
one explicit delta intermediate state have been incorporated  in the
Illinois 3NF model \cite{Pieper:2001ap}. Earlier, Coon and Friar \cite{Coon:1986kq}
 systematically constructed the $1/m$ 
(with $m$ the nucleon mass) corrections to the $2\pi$ exchange 3NF, and the
so-called drift effects due to the boost of the two-nucleon force were worked
out in \cite{Robilotta:2006xq}. We will compare to these works in our 
subsequent paper, where we discuss the shorter-ranged part of the 3NF at this
order and the corresponding $1/m$ corrections. 

This manuscript is organized as follows. In Sect.~\ref{sec:3nf} we first 
write down the effective chiral Lagrangian and discuss the general structure
of the 3NF at N$^3$LO. In the subsequent subsections \ref{sec:2PE},
\ref{sec:2PE1PE}, and \ref{sec:ring} we give our results for 
the $2\pi$, $2\pi$-$1\pi$ and ring graphs, respectively,
both in momentum and coordinate space
representations. We end with a summary and conclusions in
Sect.~\ref{sec:summary}. In particular, we make some comments on the
3NF arising in an extension of this work with explicit delta degrees of
freedom.
The lengthy expressions for the momentum space representation of the ring
diagrams are relegated to the appendix.

\section{Long-range contributions to the 3NF at N$^3$LO}
\def\theequation{\arabic{section}.\arabic{equation}}
\label{sec:3nf}

The calculations performed in the following are based on the effective
chiral Lagrangian for pions and nucleons. We employ here the
heavy baryon formulation and display only the terms of relevance for
our study:
\beqa
\label{lagr}
{\cal L} &=& {\cal L}^{\Delta = 0} + {\cal L}^{\Delta = 1} + {\cal L}^{\Delta
  = 2} +  \ldots  \nn
{\cal L}^{\Delta = 0} & = & \frac{F^2}{4} \langle \nabla^\mu U \nabla_\mu 
U^\dagger + \chi_+ \rangle + \bar{N} \left(
 i \, v\cdot D + {g}_A \, u \cdot S \right) N + \ldots \,, \nn
{\cal L}^{\Delta = 1} & = & \bar{N} \left(  c_1 \, \langle \chi_+ \rangle + c_2 \, (v \cdot
       u)^2 + c_3 \, u \cdot u + c_4 \, [ S^\mu, S^\nu ] u_\mu u_\nu \right)
  N   + \ldots \,, \nn
{\cal L}^{\Delta = 2} & = & \bar{N} \left(
d_{16} S \cdot u \langle \chi_+ \rangle + i d_{18} S^\mu [ D_\mu , \, \chi_-]
 + 
\tilde d_{28} ( i \langle \chi_+ \rangle v \cdot D + \mbox{h.c.} ) 
+ \ldots \right) N + \ldots \,,
\eeqa
where the $c_i$ are low-energy constants and 
$N$, $v_\mu$  and  $S_\mu$  denote the large component of the nucleon
field, the nucleons four-velocity and the covariant spin vector, respectively.  
We use standard notation:
$U(x) = u^2(x)$ collects the pion fields, $u_\mu = i(u^\dagger \partial_\mu u
- u \partial_\mu u^\dagger)$, $\chi_+ = u^\dagger \chi u^\dagger + u
\chi^\dagger u$ includes the explicit chiral symmetry breaking due to the
finite light quark masses,
$\langle \ldots \rangle$ denotes a trace in flavor space and $D_\mu$
is the chiral covariant derivative for the nucleon  
field. Notice further that the first terms in the expansion of 
$U(\fet \pi )$ in powers of the pion fields read:
\beq
\label{matrU}
U (\fet \pi ) = 1 + \frac{i}{F_\pi} \fet \tau \cdot \fet \pi - 
\frac{1}{2 F_\pi^2} \fet \pi^2 - \frac{i \alpha}{F_\pi^3} 
(\fet \tau \cdot \fet \pi )^3 + \frac{8 \alpha - 1}{8 F_\pi^4} \fet \pi^4 + \ldots\,,
\eeq
where $\fet \tau$ denote the Pauli isospin matrices and $\alpha$ is an
arbitrary  constant. For further notation and discussion, we refer 
to Ref.~\cite{Fettes:1998ud}, a recent review is given in Ref.~\cite{Bernard:2007zu}.
Following Weinberg \cite{Weinberg:1990rz,Weinberg:1991um}, the dimension
$\Delta$ of the Lagrangian is  defined via
\beq
\Delta = d + \frac{1}{2} n - 2 \,,
\eeq
where $d$ and $n$ are the number of derivatives or insertions of the pion mass $M_\pi$ and
nucleon field operators, respectively. 
The pertinent low-energy constants (LECs) of the leading-order effective Lagrangian 
are the nucleon axial-vector coupling ${g}_A$ and the pion decay constant
$F_\pi$. 
Notice that while all couplings and masses appearing in the effective
Lagrangian should, strictly speaking, be
taken at their SU(2) chiral limit values, to the accuracy we are working, 
we can use their pertinent physical values. 
In addition, we have the LECs 
$d_{16}$, $d_{18}$ and $\tilde d_{28}$ from the $\pi N$
Lagrangian at order $\Delta = 2$. The ellipses in the brackets in the second line of
Eq.~(\ref{lagr}) refer to terms proportional to the LECs $d_{1,2,3,5,14,15
}$ and $\tilde d_{24,26,27,28,30}$ which generate $\pi \pi NN$ vertices
\cite{Fettes:1998ud} but
do not contribute to the 3NF at N$^3$LO as will be shown later. 
We also omit in Eq.~(\ref{lagr}) pion vertices with $\Delta
  =2$ and proportional to LECs $l_{3,4}$ which do not show up explicitly
in the formulation based on renormalized pion fields at the considered order, see
  Ref.~\cite{Epelbaum:2002gb} for more details. 

For a connected $N$-nucleon diagram
with $L$ loops and $V_i$ vertices of dimension $\Delta_i$, the  
irreducible contribution\footnote{This is the contribution which is not
generated through iterations in the dynamical equation and which gives rise
to the nuclear force.} 
to the scattering amplitude scales as $Q^\nu$, where $Q$ is a generic
low-momentum scale associated with external nucleon three-momenta or $M_\pi$ and  
\beq
\nu = -4 + 2 N + 2 L + \sum_i V_i \Delta_i \,.
\eeq 
Consequently, at N$^3$LO, which corresponds to $\nu = 4$, one needs to take
into account two classes of
connected diagrams: tree diagrams with one insertion of the $\Delta
=2$-interactions
and one-loop graphs involving only lowest-order vertices with $\Delta=0$.  
Notice that it is not possible to draw 3N diagrams with two insertions of 
$\Delta =1$-vertices at this order. We further emphasize that similar to the
case of the leading four-nucleon force considered in Refs.~\cite{Epelbaum:2006eu,Epelbaum:2007us}, 
disconnected diagrams lead to vanishing contributions to the 3NF and will not
be discussed in what follows.  

\begin{figure}[tb]
\vskip 1 true cm
\includegraphics[width=15.0cm,keepaspectratio,angle=0,clip]{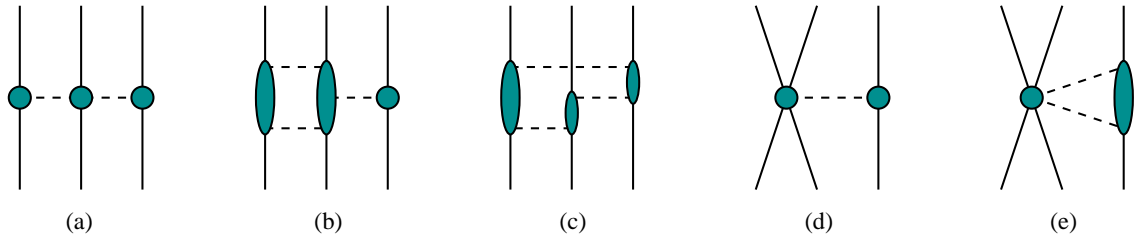}
    \caption{
         Various topologies that appear in the 3NF at N$^3$LO. Solid and dashed lines
         represent nucleons and pions, respectively. Shaded blobs are the
         corresponding amplitudes. The long-range part of the 3NF considered
         in this paper consists of (a) $2\pi$ exchange graphs, (b) $2\pi$-$1\pi$
         diagrams and the so-called ring diagrams (c). The topologies (d) and
         (e) involve four-nucleon contact operators and are considered of
         shorter range. 
\label{fig1} 
 }
\end{figure}

The structure of the 3NF at N$^3$LO is visualized in
Fig.~\ref{fig1}
and can be written as 
\beq
V_{\rm 3N}^{(4)} = V_{\rm 2\pi}^{(4)} +   V_{\rm 2\pi \mbox{-} 1\pi}^{(4)} 
 +   V_{\rm ring}^{(4)} +   V_{\rm 1\pi \mbox{-} cont}^{(4)}  +   V_{\rm 2\pi \mbox{-}
   cont}^{(4)} 
+   V_{1/m}^{(4)}  \,.
\eeq
While the $2\pi$-$1\pi$, ring and  
two-pion-exchange-contact ($2\pi$-cont) topologies start to contribute at  N$^3$LO,
the $2\pi$ and one-pion-exchange-contact
($1\pi$-cont) graphs already appear at N$^2$LO yielding the
following contributions to the 3NF \cite{vanKolck:1994yi,Epelbaum:2002vt}:
\beqa
\label{leading}
V_{\rm 2\pi}^{(3)} &=& \frac{g_A^2}{8 F_\pi^4}\; 
\frac{\vec \sigma_1 \cdot \vec q_1  \; \vec \sigma_3 \cdot \vec q_3 }{[
  q_1^2 + M_\pi^2] \, [q_3^2 + M_\pi^2]} \;\Big[ \fet \tau_1
  \cdot \fet \tau_3  \, \left( - 4 c_1 M_\pi^2 + 2 c_3 \, \vec q_1 \cdot \vec
    q_3 \right)  
+  c_4 \fet \tau_1
  \times \fet \tau_3  \cdot \fet \tau_2  \; \vec q_1 \times \vec q_3 
\cdot \vec \sigma_2  \Big]  \,, \nn
V_{\rm 1\pi \mbox{-} cont}^{(3)}   &=& - \frac{g_A \, D}{8 F_\pi^2}\;  
\frac{\vec \sigma_3 \cdot \vec q_3 }{q_3^2 + M_\pi^2} \; 
\fet \tau_1 \cdot \fet \tau_3 \; \vec \sigma_1 \cdot \vec q_3\,,
\eeqa
where the subscripts refer to the nucleon labels and $\vec q_{i} = \vec p_i \,
' - \vec p_i$,  with $\vec p_i \, '$
and $\vec p_i$ being the final and initial momenta of the nucleon $i$. 
Further, $q_i \equiv | \vec q_i |$, $\sigma_i$  denote the Pauli
spin matrices and 
$D$ refers to the low-energy constant
accompanying the 
leading 
$\pi NNNN$
vertex. Here and throughout this work, the results are always
given for a particular choice of nucleon labels. The full expression for
the 3NF results by taking into account all possible permutations of the
nucleons\footnote{For 
three nucleons there are altogether 6 permutations.}, i.e.:
\beq
V_{\rm 3N}^{\rm full} = V_{\rm 3N} + \mbox{all permutations}\,.
\eeq
In this work, we focus on
the long-range contributions $V_{\rm 2\pi}^{(4)}$, $V_{\rm 2\pi \mbox{-} 1\pi}^{(4)}$
and $V_{\rm ring}^{(4)}$
resulting from diagrams (a), (b) and (c) in Fig.~\ref{fig1}, respectively. 
The remaining terms resulting from graphs (d) and (e) in Fig.~\ref{fig1} 
and the relativistic $1/m$-corrections will be discussed 
separately.

\subsection{Two-pion exchange topology}
\label{sec:2PE}

\begin{figure}[tb]
\vskip 1 true cm
\includegraphics[width=15.0cm,keepaspectratio,angle=0,clip]{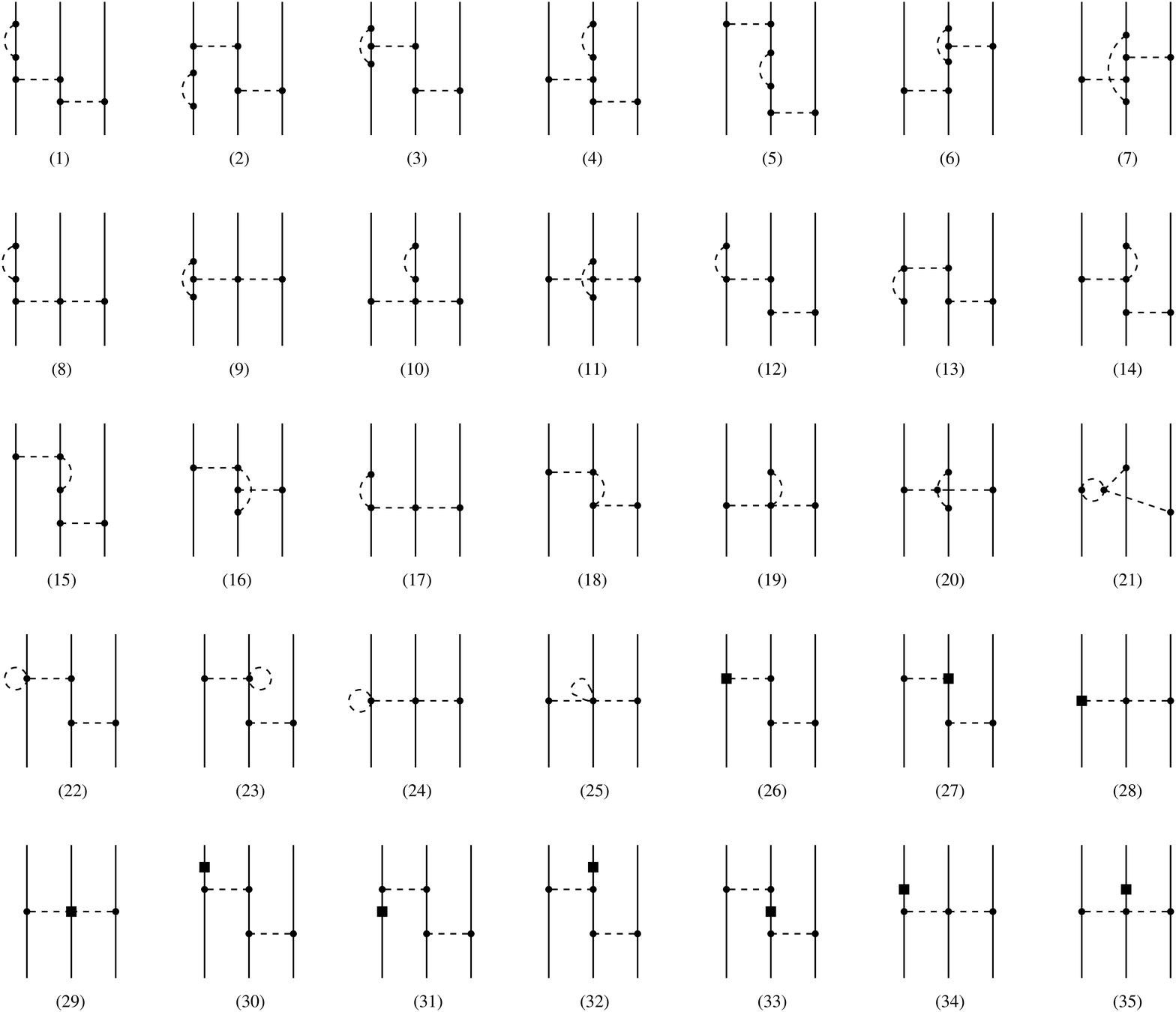}
    \caption{
         Two-pion exchange 3N diagrams at N$^3$LO. Solid dots (filled
 rectangles) denote vertices of dimension $\Delta_i = 0$ ($\Delta_i =2$). 
 Diagrams which result from the interchange of the nucleon lines and/or
 application of the time reversal operation are not shown. For remaining
 notation see Fig.~\ref{fig1}.
\label{fig2} 
 }
\end{figure}

The $2\pi$ exchange diagrams at $\nu = 4$ corresponding to topology (a) of
Fig.~\ref{fig1} are depicted in Fig.~\ref{fig2}.
We do not show graphs involving tadpoles at pion lines which just renormalize the
pion field and mass, see \cite{Epelbaum:2002gb} for details. A close inspection
of diagrams in Fig.~\ref{fig2} shows that most of them do not generate 3NFs. 
First, the contribution from graphs (12-17) involves an odd power 
of the loop momentum $\vec l$ to be integrated over and thus vanishes.
Second, diagrams (1), (2), (4), (8), (10), (30-32), (34) and (35)  
just renormalize the external nucleon legs. Similarly, Feynman diagrams  
(3), (9), (22-24), (26-28) lead to renormalization of the
leading pion-nucleon coupling without producing any new
structures. All these contributions are taken into account by replacing the
bare LECs in the leading $2\pi$ exchange 3N scattering amplitude by renormalized
ones. This suggests that there are no N$^3$LO corrections to the 3NF from
these graphs since the $2\pi$ exchange 3N diagrams at order $\nu = 2$ do not generate any 
nonvanishing 3NF. Given the fact that nuclear potentials are, in general, not uniquely
defined, the above argument based on the (on-shell) scattering amplitude should be
taken with care. We have, however, verified that this is indeed the case by 
explicitly calculating the corresponding 3NF using the method of unitary
transformation along the lines of Ref.~\cite{Epelbaum:2007us}. 
From the remaining graphs in Fig.~\ref{fig2}, diagram (11) does not contribute at the considered
order due to the $1/m$-suppression caused by the time derivative entering the
Weinberg-Tomozawa vertex.\footnote{This graph does not involve reducible
  time-ordered topologies. Its contribution to the nuclear force is,
  therefore, most easily obtained using the Feynman graph technique. The
  $1/m$-suppression from the time derivative entering the Weinberg-Tomozawa vertex
  follows then simply from the four-momentum conservation.} 
For the same reason, diagram (25) also leads to a vanishing result at the
order considered. Here, the
time derivative acts either on the pions exchanged between two nucleons 
leading to a $1/m$-suppression or on the pion in the tadpole giving an odd power 
of the loop momentum $l^0$ to be integrated over. 
Further, it is easy to see that Feynman diagrams (18) and (21) do not
contribute as well. Diagram (29)
involves one insertion of the $\pi \pi NN$ vertices of dimension $\nu
=2$. The relevant vertices are proportional to LECs $d_{1,2,3,5,14,15
}$ and $\tilde d_{24,26,27,28,30}$. 
The corresponding 3NF is shifted to higher orders since all these vertices involve
at least one time derivative, see \cite{Fettes:1998ud} for explicit
expressions. Last but not least, we also found that diagram (33) does not
generate any 3NF. 
Thus, we are left with diagrams (5-7), (19) and (20). 
The 3NF contribution from diagrams (5-7) can be evaluated straightforwardly
using the expressions for the effective Hamilton operator from 
Ref.~\cite{Epelbaum:2007us}. Diagrams (19) and (20) do not involve reducible
topologies and can be evaluated using the
Feynman graph technique. Notice that the individual
contributions from graphs (19)  and (20) in Fig.~\ref{fig2} and
from diagram (20) in Fig.~\ref{fig3} depend on the
arbitrary constant $\alpha$ which specifies the parametrization of the matrix
$U$, see Eq.~(\ref{matrU}). Clearly, their sum is $\alpha$-independent. 

We are now in the position to present our results. The expressions for 
diagrams (5-7) and (19)  can be cast into the form of Eq.~(\ref{leading})
leading only to shifts in  the values of the LECs $c_i$   
\beq
\label{corrections_c}
c_1 \to \bar c_1 = c_1 - \frac{g_A^2 \,M_\pi}{64 \pi F_\pi^2}\,, \quad  \quad 
c_3 \to \bar c_3 = c_3 + \frac{g_A^4 \,M_\pi}{16 \pi F_\pi^2}\,, \quad  \quad 
c_4 \to \bar c_4 = c_4 - \frac{g_A^4 \,M_\pi}{16 \pi F_\pi^2}\,,
\eeq
with $\delta c_1 = -0.13\,$GeV$^{-1}$ and  
$\delta c_3 = -\delta c_4 = 0.52\,$GeV$^{-1}$.
These shifts are of the order of 20\% to 30\% of the corresponding LECs, and
thus can not be neglected in precision studies of 3NFs. In contrast to this,
the contribution from graph (20) takes a more complicated form compared to 
Eq.~(\ref{leading}) and is given by
\beqa
\label{tpe3}
V_{2\pi}^{(4)} &=& \frac{g_A^4}{256 \pi F_\pi^6} \,\frac{\vec \sigma_1 \cdot \vec
  q_1  \; \vec \sigma_3 \cdot \vec q_3 }{[
  q_1^2 + M_\pi^2] \, [q_3^2 + M_\pi^2]} \;\Big[ \fet \tau_1
  \cdot \fet \tau_3  \, \big( M_\pi (M_\pi^2 + 3 q_1^2 + 3 q_3^2 + 4 \vec q_1 \cdot \vec q_3 ) + 
(2 M_\pi^2 + q_1^2 + q_3^2 + 2 \vec q_1 \cdot \vec q_3 ) \\
&& {} \times (3M_\pi^2 + 3 q_1^2 + 3 q_3^2 + 4 \vec q_1 \cdot \vec q_3 ) A (q_2) \big)
 - \fet \tau_1
  \times \fet \tau_3  \cdot \fet \tau_2  \; \vec q_1 \times \vec q_3 
\cdot \vec \sigma_2  \; 
 \big( M_\pi + ( 4 M_\pi^2 + q_1^2 + q_3^2 + 2 \vec q_1 \cdot \vec q_3 ) A
 (q_2 ) \big) \Big]\,.
\nonumber
\eeqa
Here, we have used dimensional regularization to evaluate the loop
integrals. In this framework, the loop function $A(q)$ is given by:
\beq 
A(q) = \frac{1}{2 q} \arctan \frac{q}{2 M_\pi}\,.
\eeq  
We further emphasize that the above expressions correspond to the choice $\alpha = 0$. 
Notice also that some pieces in Eq.~(\ref{tpe3}) can be brought into a form
corresponding to the $2\pi$-$1\pi$ and $2\pi$-cont topologies by canceling
out the pion propagators with terms in the numerator. 
The $2\pi$ exchange contributions arising from diagrams in Fig.~\ref{fig2} have
also been considered recently by Ishikawa and Robilotta based on the  infrared
regularization \cite{Ishikawa:2007zz}. We have verified that the long-range
$2\pi$ exchange contributions in Eqs.~(\ref{leading}), (\ref{corrections_c}) and (\ref{tpe3})
agree with the ones given in Ref.~\cite{Ishikawa:2007zz} provided the latter are
expanded in powers of $1/m$ and the leading terms are kept.

The coordinate space representation of the $2\pi$ exchange 3NF can be obtained in
a straightforward way. For the leading terms in the first line of
Eq.~(\ref{leading}) and the corrections in Eq.~(\ref{corrections_c})
one gets
\beqa
\label{temp5}
V_{\rm 2\pi} (\vec r_{12}, \, \vec r_{32} \, ) &=& \int \frac{d^3 q_1}{(2
  \pi)^3} \,  \frac{d^3 q_3}{(2 \pi)^3} \, e^{i \vec
  q_1 \cdot \vec r_{12}} \; e^{i \vec   q_3 \cdot \vec r_{32}} \;V_{\rm
  2\pi}(\vec q_{1}, \, \vec q_{3} ) \nn
&=& \frac{g_A^2 M_\pi^6}{128 \pi^2 F_\pi^4} 
\; \vec \sigma_1 \cdot \vec \nabla_{12} \; \vec \sigma_3 \cdot \vec \nabla_{32} \;
\left[ \fet \tau_1 \cdot \fet \tau_3 \, \left( 4 \bar c_1  + 2 \bar
    c_3  \vec \nabla_{12} \cdot \vec \nabla_{32} \right) + \bar c_4
\fet \tau_1 \times \fet \tau_3 \cdot \fet \tau_2 
\; \vec \nabla_{12} \times \vec \nabla_{32} \cdot \vec \sigma_2 \right]
\nn
&& {} \times \, U_1
(x_{12} ) \; U_1(x_{32} ) \,,
\eeqa
where $\vec r_{ij} \equiv \vec r_i - \vec r_j$ is the distance between the
nucleons $i$ and $j$, the  $\vec x_i \equiv M_\pi \, \vec r_i$ are
dimensionless distances, the $\vec \nabla_i$ act on $\vec x_i$ and 
$x_{ij} \equiv | \vec x_{ij}  |$. Further, the scalar function $U_{1} (x)$ is
defined as:
\beq
U_1 (x) = \frac{4 \pi }{M_\pi} \int \frac{d^3 q}{(2 \pi )^3} \, \frac{e^{i
    \vec q \cdot \vec x/M_\pi}}{q^2 + M_\pi^2} = \frac{e^{-x}}{x}\,.
\eeq
Similarly, one obtains for the terms in
Eq.~(\ref{tpe3}) which do not involve the loop function $A(q_2)$:
  \beqa
\label{temp6}
V_{\rm 2\pi} (\vec r_{12}, \, \vec r_{32} \, ) &=& \frac{g_A^4 M_\pi^7}{4096 \pi^3 F_\pi^6}
\, \vec \sigma_1 \cdot \vec \nabla_{12} \; \vec \sigma_3 \cdot \vec \nabla_{32} \; 
\Big[ \fet \tau_1 \cdot \fet \tau_3 \, \left( - 1 + 3\nabla_{12}^2 +
  3\nabla_{32}^2  + 4 \vec  \nabla_{12}  \cdot \vec \nabla_{32}  
\right) \nn
&& {} - \fet \tau_1 \times \fet \tau_3 \cdot \fet \tau_2 
\; \vec \nabla_{12} \times \vec \nabla_{32} \cdot \vec \sigma_2 \Big] \, U_1
(x_{12} ) \; U_1(x_{32} ) \,.
\eeqa
It should be understood that the obtained expressions are only valid in the
region of space where the interparticle distances are large (i.e.~larger than
the inverse pion mass). The behavior of the potential at shorter distances is,
in general, affected by the regularization procedure which is not considered in
the present work. 

To obtain the coordinate space representation for the terms in Eq.~(\ref{tpe3})
which involve the loop function $A(q_2)$, it
is more convenient to proceed in a different way in order to avoid a
complicated angular integration:
\beqa
\label{temp1}
V_{\rm 2\pi} (\vec r_{12}, \, \vec r_{32} \, )  &=& \int \frac{d^3 q_1}{(2 \pi)^3}
\, \frac{d^3 q_2}{(2 \pi)^3} \, \frac{d^3 q_3}{(2 \pi)^3} \, 
(2 \pi )^3 \delta^3 (\vec q_1 + \vec q_2 + \vec q_3 \,) \, 
e^{i \vec
  q_1 \cdot \vec r_{1}} \; e^{i \vec
  q_2 \cdot \vec r_{2}} \; e^{i \vec   q_3 \cdot \vec r_{3}} \;V_{\rm
  2\pi}(\vec q_{1}, \, \vec q_{2}, \, \vec q_3) \nn
&=& \int d^3 r_0 \, \int \frac{d^3 q_1}{(2 \pi)^3}
\, \frac{d^3 q_2}{(2 \pi)^3} \, \frac{d^3 q_3}{(2 \pi)^3} \, 
e^{i \vec
  q_1 \cdot \vec r_{10}} \; e^{i \vec
  q_2 \cdot \vec r_{20}} \; e^{i \vec   q_3 \cdot \vec r_{30}} \;V_{\rm
  2\pi}(\vec q_{1}, \, \vec q_{2}, \, \vec q_{3} ) \nn
&=& -
\frac{g_A^4 M_\pi^7}{4096 \pi^3 F_\pi^6}
\, \vec \sigma_1 \cdot \vec \nabla_{12} \; \vec \sigma_3 \cdot \vec \nabla_{32} \; 
\Big[ \fet \tau_1 \cdot \fet \tau_3 \, 
(2 - \nabla_{12}^2 -  \nabla_{32}^2 - 2 \vec \nabla_{12} \cdot \vec \nabla_{32} )
(3 - 3\nabla_{12}^2 -  3\nabla_{32}^2 - 4 \vec \nabla_{12} \cdot \vec
\nabla_{32} ) \nn
&& {} + \fet \tau_1 \times \fet \tau_3 \cdot \fet \tau_2 \; \vec \nabla_{12}
\times \vec \nabla_{32} \cdot \vec \sigma_2 \; (4 - \nabla_{12}^2 -
\nabla_{32}^2 - 2 \vec \nabla_{12} \cdot \vec \nabla_{32} ) \Big] \nn
&& {} \times \frac{1}{4 \pi}
\int d^3 x \, 
U_1 (|\vec x_{12} + \vec x \,| ) \; W_1 (x) \; U_1 (|\vec x_{32} + \vec x \,| ) 
 \,,
\eeqa
where 
\beq
W_1(x) = \frac{4 \pi }{M_\pi^2} \int \frac{d^3 q}{(2 \pi )^3} \, e^{i
    \vec q \cdot \vec x/M_\pi}\, A (q)  = \frac{e^{-2 x}}{2 x^2}\,.
\eeq
For various techniques to evaluate the integral in the last line of Eq.~(\ref{temp1})
the reader is referred to Ref.~\cite{Ishikawa:2007zz}.

\subsection{Two-pion--one-pion exchange topology}
\label{sec:2PE1PE}

\begin{figure}[tb]
\vskip 1 true cm
\includegraphics[width=15.0cm,keepaspectratio,angle=0,clip]{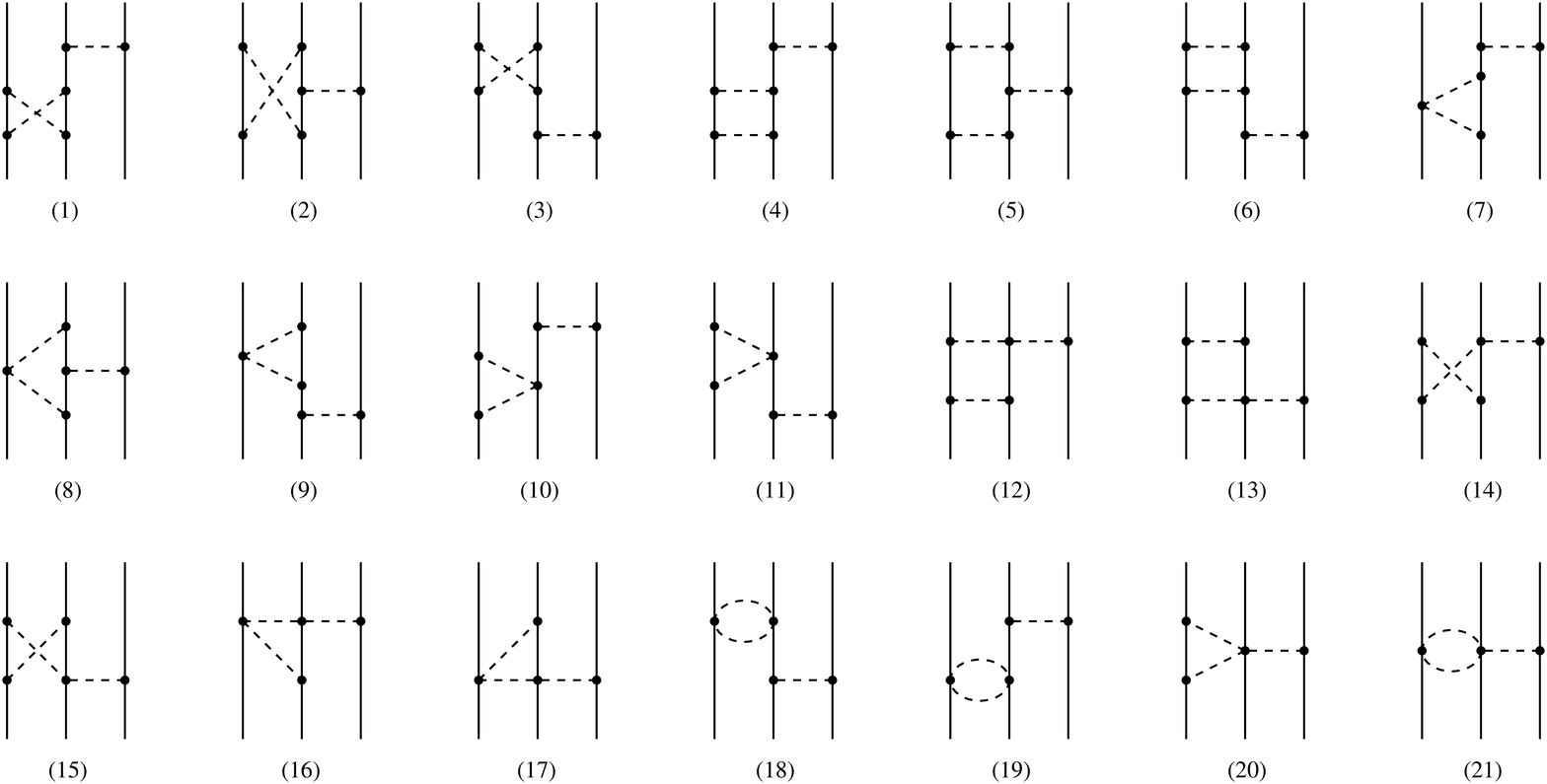}
    \caption{
         $2\pi$-$1\pi$ diagrams at N$^3$LO. Graphs resulting
         from the interchange of the nucleon lines are not shown. For notation
         see Figs.~\ref{fig1}, \ref{fig2}.
\label{fig3} 
 }
\end{figure}

Consider now the $2\pi$-$1\pi$ 3NF arising from diagrams shown in
Fig.~\ref{fig3} corresponding to topology (b) in Fig.~\ref{fig1}.
They can be written in the form:
\beqa
V_{\rm 2\pi \mbox{-}1\pi} &=& 
 \frac{\vec \sigma_3 \cdot \vec q_3}{q_3^2 + M_\pi^2} \Big[ \fet \tau_1
  \cdot \fet \tau_3  \; \left( \vec \sigma_2 \cdot \vec q_1 \; \vec q_1 \cdot
    \vec q_3 \; F_1 (q_1)  + \vec \sigma_2 \cdot \vec q_1 \; F_2 (q_1) + 
  \vec \sigma_2 \cdot \vec q_3 \;  F_3 (q_1)  \right) + \fet \tau_2
  \cdot \fet \tau_3 \; (  \vec \sigma_1 \cdot \vec q_1 \;  \vec q_1 \cdot
    \vec q_3 \; F_4 (q_1) \nn
&& {} +  \vec \sigma_1 \cdot \vec q_3 \; F_5  (q_1) +
\vec \sigma_2 \cdot \vec q_1 \;  F_6 (q_1) + 
  \vec \sigma_2 \cdot \vec q_3 \;   F_{7} (q_1)
)  + \fet \tau_1
  \times \fet \tau_2 \cdot \fet \tau_3 \; \vec \sigma_1 \times \vec \sigma_2
  \cdot \vec q_1 \;  F_{8} (q_1)
\Big]\,,
\eeqa
with $F_{1, \ldots ,8} (q_1) $ being scalar functions.  
Using the formal expressions for the effective Hamiltonian
from Ref.~\cite{Epelbaum:2007us} we obtain the following result for the first six graphs:
\beqa
F_1 (q_1) &=& - \frac{g_A^6}{256 \pi F_\pi^6} \;
\bigg[
\frac{M_\pi}{4 M_\pi^2 + q_1^2} + \frac{2 M_\pi}{q_1^2} - 
\frac{8 M_\pi^2 + q_1^2}{q_1^2} A (q_1) \bigg] \,, \nn
F_3 (q_1) &=& -\frac{g_A^6}{256 \pi F_\pi^6} \;
\Big[3 M_\pi + ( 8 M_\pi^2 + 3 q_1^2)
A(q_1) \Big] \,, \nn
F_4 (q_1) &=& - \frac{1}{q_1^2} F_5 (q_1) =
- \frac{g_A^6}{128 \pi F_\pi^6} \;
A(q_1) \,. 
\eeqa

Consider now diagrams (7-15) which involve one insertion of the
Weinberg-Tomozawa vertex. The first three graphs only 
contribute to the functions $F_{1,3} (q_1)$:
\beqa
F_1 (q_1) &=& \frac{g_A^4}{256 \pi F_\pi^6} \;
\bigg[
\frac{ M_\pi}{q_1^2} + \frac{q_1^2 - 4 M_\pi^2 }{q_1^2} \, A (q_1 ) \bigg] \,, \nn
F_3 (q_1 )&=& \frac{g_A^4}{256 \pi F_\pi^6} \;
\Big[M_\pi + (q_1^2 + 4 M_\pi^2 ) \, A (q_1 ) \Big]\,.
\eeqa
We find that diagrams (10) and (11) do not generate a 3NF while 
the contribution from graphs (12-15) has the form:
\beq
F_2 (q_1 )= \frac{g_A^4}{128 \pi F_\pi^6} \;
\Big[ M_\pi + (q_1^2 + 2 M_\pi^2 ) \, A (q_1 ) \Big]\,.
\eeq
The next two diagrams (16) and (17) do not involve reducible topologies and
can be dealt with using the Feynman graph technique. It is easy to see that
their contribution to the scattering amplitude is suppressed by $1/m$ due to
the time derivative which enters the Weinberg-Tomozawa vertices.  The
next two diagrams (18) and (19) yield  vanishing contributions to the 3NF
for exactly the same reason as do the two-pion exchange diagrams $\propto
g_A^4$ at order $\nu =
2$. Next, the contribution from diagram (20) reads: 
\beqa
F_6 (q_1 )&=& 2 F_{7} (q_1 )= \frac{g_A^4}{64 \pi F_\pi^6} \;
\Big[ M_\pi + (q_1^2 + 2 M_\pi^2 ) \, A (q_1 ) \Big] \,, \nn
F_{8} (q_1) &=& - \frac{g_A^4}{512 \pi F_\pi^6} \;
\Big[ M_\pi + (q_1^2 + 4 M_\pi^2 ) \, A (q_1 ) \Big]\,.
\eeqa
Finally, it is easy to see that the last Feynman diagram (21) in Fig.~\ref{fig3}
yields a vanishing result at the order considered. 

The coordinate space representation of the $2\pi$-$1\pi$ 3NF can be obtained
straightforwardly employing the same Fourier-type integrations as in the
the first line of Eq.~(\ref{temp1}). This leads to the following expression
where short-range terms resulting from constant contributions to
$F_{2,3,6,7,8}$, which are proportional to $M_\pi$, are not shown:
:
\beqa
\label{coordW}
V_{\rm 2\pi \mbox{-}1\pi} (\vec r_{12}, \, \vec r_{32} \, ) &=& \frac{g_A^4
  M_\pi^7}{8192 \, \pi^3 F_\pi^6}
\,  \vec \sigma_3 \cdot \vec \nabla_{32} \; 
\Big( 2 \fet \tau_1 \cdot \fet \tau_3 \, \Big[
\vec \sigma_2 \cdot \vec \nabla_{12} \; \vec \nabla_{12} \cdot \vec
\nabla_{32} 
\big(  - 2 g_A^2 U_1 (2 x_{12} ) + (1 + g_A^2) W_1 (x_{12} )   \nn
&& {} 
+ (1 - 2 g_A^2) W_3 (x_{12} )
  \big) 
-  2 \vec \sigma_2 \cdot \vec \nabla_{12}  
\big( 2 W_1 (x_{12} )+ W_2 (x_{12} ) \big) 
- \vec \sigma_2 \cdot \vec \nabla_{32} 
\big( 4 (1 - 2 g_A^2) W_1 (x_{12} ) \nn
&& {} + (1 - 3 g_A^2) W_2 (x_{12} ) \big) \Big]
- 4 \fet \tau_2 \cdot \fet \tau_3 \, \Big[
g_A^2 \; \vec \sigma_1 \cdot \vec \nabla_{12} \; \vec \nabla_{12} \cdot \vec
\nabla_{32} \; W_1 (x_{12} ) 
+ g_A^2 \;  \vec \sigma_1 \cdot \vec \nabla_{32}  \; W_2 (x_{12} )  \nn
&& {} + 2 \;  \vec \sigma_2 \cdot \vec \nabla_{12}  \; \big( 2 W_1 (x_{12} )
+ W_2 (x_{12} ) \big) 
+   \vec \sigma_2 \cdot \vec \nabla_{32}  \; \big( 2 W_1 (x_{12} )
+ W_2 (x_{12} ) \big) \Big] \nn
&& {} 
 + \fet \tau_1
  \times \fet \tau_2 \cdot \fet \tau_3 \; \vec \sigma_1 \times \vec \sigma_2
  \cdot \vec \nabla_{12} \; \big( 4 W_1 (x_{12} )
+ W_2 (x_{12} ) \big) \Big) \, U_1 (x_{32}) \,,
\eeqa
with
\beqa
W_2 (x) &=& - \nabla_x^2  W_1 (x) = - \frac{e^{-2 x}}{x^4} (1 + 2 x (1 + x))
\,, \nn
W_3 (x) &=&  \frac{4 \pi }{M_\pi^2} \int \frac{d^3 q}{(2 \pi )^3} \; e^{i
    \vec q \cdot \vec x/M_\pi}\, \left[ \frac{M_\pi}{q^2} - \frac{4
      M_\pi^2}{q^2} A (q)  \right]  = 2 Ei (-2 x) + \frac{e^{-2 x}}{x}\,,
\eeqa
and
\beq
Ei (x) \equiv - \int_{-x}^\infty \frac{e^{-t} \, dt}{t}\,.
\eeq
We emphasize again that the above expressions are only valid at large
distances. 

\subsection{Ring diagrams}
\label{sec:ring}

\begin{figure}[tb]
\vskip 1 true cm
\includegraphics[width=17.5cm,keepaspectratio,angle=0,clip]{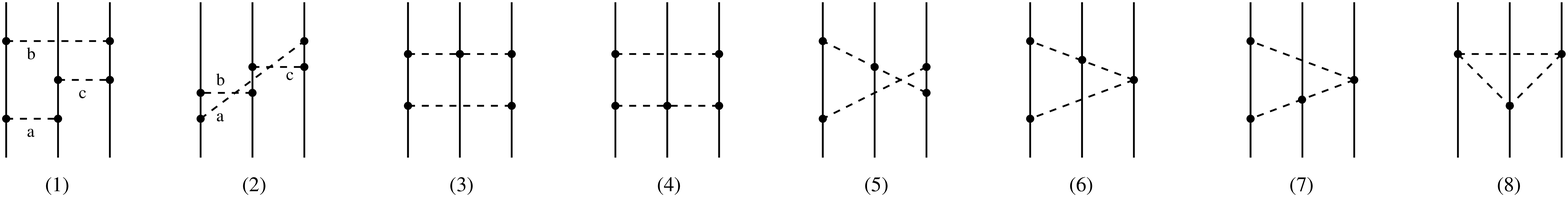}
    \caption{
         Ring diagrams at N$^3$LO. Graphs resulting
         from the interchange of the nucleon lines are not shown. For notation
         see Figs.~\ref{fig1}, \ref{fig2}.
\label{fig4} 
 }
\end{figure}

We now regard ring diagrams shown in Fig.~\ref{fig4} which correspond to topology
(c) in Fig.~\ref{fig1}. These are most cumbersome to evaluate. 
The contributions from the first two diagrams can be obtained using the
expressions for the effective Hamilton operator given in Ref.~\cite{Epelbaum:2007us}. 
This leads to the following structures:
\beqa
V_{\rm ring}^1 &=& M^1 \left[ \frac{4}{\omega_a^3 \, \omega_b \, \omega_c} +   
\frac{4}{\omega_a \, \omega_b^3 \, \omega_c} - \frac{4}{\omega_a \, \omega_b \, 
  \omega_c^3} \right] \,, \nn
V_{\rm ring}^2 &=& M^2 \left[ - \frac{4}{\omega_a^3  \, \omega_b  \, \omega_c} -   
\frac{4}{\omega_a \,  \omega_b^3  \, \omega_c} - \frac{4}{\omega_a  \, \omega_b \, 
  \omega_c^3} \right]\,, 
\eeqa
where $M^i$ represents the spin, isospin and momentum structure which results
from the vertices entering the diagram $i$ and $\omega$ denotes the pion free
energy, $\omega \equiv \sqrt{q^2 + M_\pi^2}$. 
Substituting the expressions for the vertices, the result can be written in the
form 
\beq
\label{ring:temp}
V_{\rm ring} = \left( \frac{g_A}{2 F_\pi} \right)^6 \frac{1}{(2 \pi)^3}
\int \, d^3l_1 \, d^3l_2 \, d^3l_3 \,  \delta^3 (\vec l_3 - \vec l_2 - \vec
q_1 ) \, \delta^3 (\vec l_2 - \vec l_1 - \vec q_3 ) 
\frac{v}{[l_1^2 + M_\pi^2] \, [l_2^2 + M_\pi^2]^2 \, [l_3^2 + M_\pi^2]}
 \,,
\eeq
with the numerator 
\beqa
v &=& - 8 \fet \tau_1 \cdot \fet \tau_2 \;  \vec l_1 \times \vec l_3  \cdot
\vec \sigma_2\;  \vec l_1 \times \vec l_2  \cdot \vec \sigma_3 \; \vec l_2
\cdot \vec l_3 \; - \;  4 \fet \tau_1 \cdot \fet \tau_3 \; 
\vec l_1 \cdot \vec l_2 \; \vec l_1 \cdot \vec l_3 \; \vec l_2 \cdot \vec l_3
\;  + \; 2  \fet \tau_1 \times \fet \tau_2  \cdot \fet \tau_3 
\;  \vec l_1 \times \vec l_3  \cdot \vec \sigma_2 \;
\vec l_1 \cdot \vec l_2 \; \vec l_2 \cdot \vec l_3 \nn
&& {}  +  
6  \vec l_2 \times \vec l_3  \cdot
\vec \sigma_1\;  \vec l_1 \times \vec l_2  \cdot \vec \sigma_3 \; \vec l_1
\cdot \vec l_3 \,.
\eeqa
Carrying out the two trivial integrations over, say, $l_1$ and $l_2$ leads to
the standard three-point function integrals.  The latter can be evaluated but the
resulting expressions are rather involved, see appendix \ref{app2}. 
It is more convenient to evaluate
Eq.~(\ref{ring:temp}) in configuration space using again the same
definition as in the the first line of Eq.~(\ref{temp5}).
This leads to the following compact result:   
\beqa
\label{ring1}
V_{\rm ring} (\vec r_{12}, \, \vec r_{32}\, )  &=& \left( \frac{g_A}{2 F_\pi} \right)^6 \int \, 
\frac{d^3 l_1}{(2 \pi )^3} \, \frac{d^3 l_2}{(2 \pi )^3} \, \frac{d^3 l_3}{(2 \pi )^3} \; 
e^{i \vec l_1 \cdot \vec r_{23}} \; e^{i \vec l_2 \cdot \vec r_{31}} \; 
e^{i \vec l_3 \cdot \vec r_{12}} \; \frac{v}{[l_1^2 + M_\pi^2] \, [l_2^2 +
  M_\pi^2]^2 \, 
[l_3^2 + M_\pi^2]}
\nn
&=&  - \frac{g_A^6 \, M_\pi^7}{4096\,  \pi^3 \, F_\pi^6} \Big[
-4 \fet \tau_1 \cdot \fet \tau_2 \; \vec \nabla_{23} \times \vec \nabla_{12}
\cdot \vec \sigma_2 \;  \vec \nabla_{23} \times \vec \nabla_{31}
\cdot \vec \sigma_3 \; \vec \nabla_{31} \cdot \vec \nabla_{12} \nn 
&& {} - \; 2 \fet \tau_1 \cdot \fet \tau_3 \;   \vec \nabla_{23} \cdot \vec \nabla_{31}
\; \vec \nabla_{23} \cdot \vec \nabla_{12} \;  \vec \nabla_{31} \cdot \vec
\nabla_{12}  
\;  + \; \fet \tau_1 \times \fet \tau_2 \cdot \fet
\tau_3 \; \vec \nabla_{23} \times \vec \nabla_{12}
\cdot \vec \sigma_2 \; \vec \nabla_{23} \cdot \vec \nabla_{31} \; 
\vec \nabla_{31} \cdot \vec \nabla_{12}  \nn 
&& {}  + \; 3 \vec \nabla_{31} \times \vec \nabla_{12}
\cdot \vec \sigma_1 \;  \vec \nabla_{23} \times \vec \nabla_{31}
\cdot \vec \sigma_3 \; \vec \nabla_{23} \cdot \vec \nabla_{12} \Big] \;
U_1 (x_{23}) \; U_2 (x_{31}) \; U_1 (x_{12}) \,,
\eeqa
where the derivatives should be evaluated as if the variables $\vec x_{12}$,
$\vec x_{23}$ and $\vec x_{31}$ were independent\footnote{Clearly, the
  relative distances $\vec r_{12}$, $\vec r_{23}$ and $\vec r_{31}$ are
  related via $\vec r_{12} + \vec r_{23}+ \vec r_{31} =0$.} 
and we have introduced
\beq
U_2 (x) = 8 \pi M_\pi \,  \int \frac{d^3 q}{(2 \pi )^3} \, \frac{e^{i
    \vec q \cdot \vec x/M_\pi}}{[q^2 + M_\pi^2]^2} = e^{-x}\,.
\eeq

Consider now diagrams (3-5) in Fig.~\ref{fig3} which involve one insertion of
the Weinberg-Tomozawa vertex. The corresponding contribution to the 3NF can be
evaluated in a similar way as described above and using the explicit
expressions for the effective Hamilton operator from Ref.~\cite{Epelbaum:2007us}. We
find that graphs (3) and (4) lead to vanishing 3NF while the contribution from
diagram (5) reads:
\beqa
\label{ring2}
V_{\rm ring} (\vec r_{12}, \, \vec r_{32}\, )  &=& \frac{g_A^4 \, M_\pi^7}{2048\,  \pi^3 \, F_\pi^6} \Big[
2  \fet \tau_1 \cdot \fet \tau_2 \, 
\big( \vec \nabla_{23} \cdot \vec \nabla_{31} \; \vec \nabla_{31} \cdot \vec \nabla_{12}
-  \vec \nabla_{31} \times \vec \nabla_{12}  \cdot \vec \sigma_1 \; 
 \vec \nabla_{23} \times \vec \nabla_{31}  \cdot \vec \sigma_3 \big) \nn
&& {} +  \fet \tau_1 \times \fet \tau_2 \cdot \fet \tau_3 
\; \vec \nabla_{31} \times \vec \nabla_{12} \cdot \vec \sigma_1 \; \vec
\nabla_{23} \cdot \vec \nabla_{31} \Big] 
U_1 (x_{23}) \; U_1 (x_{31}) \; U_1 (x_{12}) 
\,. 
\eeqa
Finally, we found that the contributions from the last three diagrams in
Fig.~\ref{fig3} are suppressed by the factor $1/m$ and, therefore, need not to
be taken into account at the order considered.

\section{Summary and conclusions}
\def\theequation{\arabic{section}.\arabic{equation}}
\label{sec:summary}

In this work, we have derived the long-range contributions of the 3NF in 
chiral effective field theory at N$^3$LO in the chiral expansion. These
results constitute the first systematic corrections to the leading $2\pi$
exchange that appears at N$^2$LO in the Weinberg counting. We have shown
that there exists three different topologies that contribute to the long-range
components of the chiral 3NF. While some of these simply renormalize the 
LECs of the leading order $2\pi$ exchange topology, there are many new
spin-isospin structures generated at N$^3$LO from all three topologies. 
We have given the momentum and coordinate space representations of the
various contributions. At this order, there is an arbitrariness in choosing
the multi-pion interactions, cf. Eq.~(\ref{matrU}), which allows to shuffle
strength between the various contributions. The sum of all diagrams
contributing to the 3NF is, as demanded by field theory, independent of the
choice of this parameter. To be precise, we have given our results for the
special choice $\alpha = 0$. It is important to stress that the corrections
to the 3NF discussed here are free of any unknown low-energy constants and
can be expressed entirely in terms of the nucleon axial-vector coupling $g_A$,
the pion decay constant $F_\pi$
and the pion mass $M_\pi$. It should further be emphasized that the dimension-two LECs 
$c_3$ and $c_4$ which enter the leading 3NF and would also contribute at
N$^4$LO are larger than their natural values $c_i \sim g_A/(4\pi F_\pi) \sim 1$ GeV$^{-1}$ 
\cite{Bernard:2007zu}. This  is understood
in terms of $\Delta (1232)$ $s$-channel and vector-meson $t$-channel excitations
\cite{Bernard:1996gq}. It is, therefore, expected that the theory with explicit
spin-3/2 degrees of freedom leads to an improved convergence as compared to
the pure pion-nucleon theory considered here. Of course, the theory with
explicit deltas leads to many more structures and has also been much less
developed phenomenologically. Still, it is important to confront the results
obtained here with similarly accurate calculations in the delta-full theory.
Work along these lines is in progress.

\section*{Acknowledgments}

The work of E.E. and H.K. was supported in parts by funds provided from the 
Helmholtz Association to the young investigator group  
``Few-Nucleon Systems in Chiral Effective Field Theory'' (grant  VH-NG-222)
and through the virtual institute ``Spin and strong QCD'' (grant VH-VI-231). 
This work was further supported by the DFG (SFB/TR 16 ``Subnuclear Structure
of Matter'') and by the EU Integrated Infrastructure Initiative Hadron
Physics Project under contract number RII3-CT-2004-506078.

\appendix

\def\theequation{\Alph{section}.\arabic{equation}}
\setcounter{equation}{0}
\section{Expressions for ring diagrams in momentum-space}
\label{app2}

In this appendix we give lengthy expressions for ring
diagrams in Fig.~\ref{fig4} in momentum space. The contributions from diagrams
(1) and (2) can be expressed as:
\beqa
\label{ringR}
V_{\rm ring}&=&
\vec{\sigma}_1\cdot\vec{\sigma}_2 \; {\fet \tau}_2\cdot{\fet\tau}_3  \; R_1+
\vec{\sigma}_1\cdot\vec{q}_1\vec{\sigma}_2\cdot\vec{q}_1 \; {\fet
  \tau}_2\cdot{\fet\tau}_3  \; R_2+
\vec{\sigma}_1\cdot\vec{q}_1\vec{\sigma}_2\cdot\vec{q}_3 \; {\fet
  \tau}_2\cdot{\fet\tau}_3  \; R_3+
\vec{\sigma}_1\cdot\vec{q}_3\vec{\sigma}_2\cdot\vec{q}_1 \; {\fet
  \tau}_2\cdot{\fet\tau}_3  \; R_4\nonumber\\
&+&\vec{\sigma}_1\cdot\vec{q}_3\vec{\sigma}_2\cdot\vec{q}_3 \; {\fet
  \tau}_2\cdot{\fet\tau}_3  \; R_5+{\fet\tau}_1\cdot{\fet\tau}_3 \;  R_6
+\vec{\sigma}_1\cdot\vec{q}_1\vec{\sigma}_3\cdot\vec{q}_1  \; R_7
+\vec{\sigma}_1\cdot\vec{q}_1\vec{\sigma}_3\cdot\vec{q}_3  \; R_8
+\vec{\sigma}_1\cdot\vec{q}_3\vec{\sigma}_3\cdot\vec{q}_1  \; R_9\\
&+&\vec{\sigma}_1\cdot\vec{\sigma}_3  \; R_{10}
+\vec{q}_1\cdot \vec{q}_3\times\vec{\sigma}_2 \; 
{\fet\tau}_1\cdot{\fet\tau}_2\times{\fet\tau}_3 \;  R_{11}.\nonumber
\eeqa
where the functions $R_i \equiv R_i (q_1, \, q_3,\, z)$ with $z=\hat{q}_1\cdot\hat{q}_3$
are defined as follows: 
\beqa
R_1&=&\frac{\left(-1+z^2\right) g_A^6 M_{\pi } \left(2 M_{\pi }^2+q_3^2\right) \left(q_2^2 q_3+4 M_{\pi }^2 \left(z q_1+q_3\right)\right)}{128
F^6 \pi  \left(4 \left(-1+z^2\right) M_{\pi }^2-q_2^2\right) \left(4 M_{\pi }^2 q_3+q_3^3\right)}-\frac{A\left(q_2\right) g_A^6 q_2^2 \left(2 M_{\pi
}^2 \left(q_1+z q_3\right)+z q_3 \left(-q_1^2+q_3^2\right)\right)}{128 F^6 \pi  \left(-1+z^2\right) q_1 q_3^2}-\nn
&&\frac{A\left(q_3\right) g_A^6 \left(z q_2^2 \left(z q_1-q_3\right) q_3+2 M_{\pi }^2 \left(z \left(-2+z^2\right) q_1^2-\left(1+z^2\right) q_1
q_3-z q_3^2\right)\right)}{128 F^6 \pi  \left(-1+z^2\right) q_1 q_3}+\nn
&&\frac{A\left(q_1\right) g_A^6 \left(2 M_{\pi }^2 q_2^2+q_3 \left(-z q_1^3+\left(2-3 z^2\right) q_1^2 q_3-z \left(-2+z^2\right) q_1 q_3^2+q_3^3\right)\right)}{128
F^6 \pi  \left(-1+z^2\right) q_3^2}-\nn
&&\frac{I(4:0,-q_1,q_3;0) g_A^6 q_2^2}{32 F^6 \left(-1+z^2\right) \left(4 \left(-1+z^2\right) M_{\pi
      }^2-q_2^2\right) q_3}\left(8 \left(-1+z^2\right) M_{\pi }^4
  \left(2 z q_1+\left(1+z^2\right) q_3\right)+q_2^2 q_3 \left(z^2
    q_1^2+\right.\right.\nn
&&\left.\left.z \left(-1+z^2\right) q_1 q_3-q_3^2\right)+2 M_{\pi }^2 \left(z \left(-2+z^2\right) q_1^3-\left(1+2 z^2\right) q_1^2 q_3+3 z \left(-2+z^2\right) q_1 q_3^2+\left(-3+2
z^4\right) q_3^3\right)\right) \,, \nn [5pt]
R_2&=&\frac{A\left(q_2\right) g_A^6 q_2^2 \left(-2 M_{\pi
}^2 \left(\left(1+z^2\right) q_1+2 z q_3\right)+z q_3 \left(\left(1+z^2\right) q_1^2-2 q_3^2\right)\right)}{128 F^6 \pi  \left(-1+z^2\right)^2 q_1^3
q_3^2}+\nn
&&\frac{A\left(q_3\right) g_A^6 \left(M_{\pi }^2 \left(2 z q_1^2+\left(1+3 z^2\right) q_1 q_3+2 z q_3^2\right)+z q_3 \left(-z q_1^3-z^2 q_1^2
q_3+z q_1 q_3^2+q_3^3\right)\right)}{64 F^6 \pi  \left(-1+z^2\right)^2 q_1^3 q_3}+\nn
&&\frac{A\left(q_1\right) g_A^6}{128 F^6 \pi  \left(-1+z^2\right)^2 q_1^2 q_3^2}\left(2 M_{\pi }^2 \left(\left(1+z^2\right) q_1^2+z \left(3+z^2\right) q_1 q_3+\left(1+z^2\right) q_3^2\right)+\right.\nn
&&\left.q_3 \left(-\left(z+z^3\right) q_1^3+\left(2-5 z^2+z^4\right) q_1^2 q_3+z \left(1+z^2\right) q_1 q_3^2+\left(1+z^2\right) q_3^3\right)\right)-\nn
&&\frac{I(4:0,-q_1,q_3;0) g_A^6}{32 F^6 \left(-1+z^2\right)^2 q_1^2 \left(-4 \left(-1+z^2\right)
      M_{\pi }^2+q_2^2\right) q_3}\left(q_2^4 q_3 \left(-2 z^2 q_1^2+\left(1+z^2\right) q_3^2\right)-\right.\nn
&&8 (-1+z) (1+z) M_{\pi }^4 \left(z \left(2+z^2\right) q_1^3+\left(1+2 z^2\right)^2 q_1^2 q_3+z \left(2+7 z^2\right) q_1 q_3^2+\left(1+2 z^2\right)
q_3^3\right)+\nn
&&\left.2 M_{\pi }^2 q_2^2 \left(2 z q_1^3+\left(1-z^2+6 z^4\right) q_1^2 q_3-2 z \left(-1-3 z^2+z^4\right) q_1 q_3^2+\left(3+3 z^2-4
z^4\right) q_3^3\right)\right)\nn
&&+\frac{g_A^6 M_{\pi } \left(2 M_{\pi }^2+q_3^2\right) \left(q_2^2 q_3+4 M_{\pi }^2 \left(z q_1+q_3\right)\right)}{128 F^6 \pi
 q_1^2 \left(4 \left(-1+z^2\right) M_{\pi }^2-q_2^2\right) \left(4 M_{\pi }^2 q_3+q_3^3\right)}\,, \nn [5pt]
R_3&=&-\frac{z A\left(q_2\right) g_A^6 q_2^2 \left(-4 M_{\pi }^2 \left(q_1+z q_3\right)+q_3 \left(2 z q_1^2+\left(-1+z^2\right) q_1 q_3-2 z q_3^2\right)\right)}{128
F^6 \pi  \left(-1+z^2\right)^2 q_1^2 q_3^3}-\nn
&&\frac{z A\left(q_3\right) g_A^6 }{128 F^6 \pi  \left(-1+z^2\right)^2 q_1^2 q_3^2}\left(M_{\pi }^2 \left(-2 z \left(-3+z^2\right) q_1^2+4 \left(1+z^2\right) q_1 q_3+4 z q_3^2\right)+q_3 \left(-\left(1+z^2\right)
q_1^3-\right.\right.\nn
&&\left.\left.2 z^3 q_1^2 q_3+\left(1+z^2\right) q_1 q_3^2+2 z q_3^3\right)\right)-\nn
&&\frac{z A\left(q_1\right) g_A^6 \left(2 M_{\pi }^2 \left(2 q_1^2+4 z q_1 q_3+\left(1+z^2\right) q_3^2\right)+q_3 \left(-2 z q_1^3+\left(1-3
z^2\right) q_1^2 q_3+2 z q_1 q_3^2+\left(1+z^2\right) q_3^3\right)\right)}{128 F^6 \pi  \left(-1+z^2\right)^2 q_1 q_3^3}-\nn
&&\frac{I(4:0,-q_1,q_3;0) z g_A^6}{32 F^6 \left(-1+z^2\right)^2 q_1 \left(-4 \left(-1+z^2\right)
      M_{\pi }^2+q_2^2\right) q_3^2}\left(q_2^4 q_3 \left(\left(1+z^2\right) q_1^2+z \left(-1+z^2\right) q_1 q_3-\left(1+z^2\right) q_3^2\right)+\right.\nn
&&8 (-1+z) (1+z) M_{\pi }^4 \left(3 z q_1^3+\left(-1+10 z^2\right) q_1^2 q_3+3 z \left(1+2 z^2\right) q_1 q_3^2+\left(1+2 z^2\right) q_3^3\right)+\nn
&&\left.2 M_{\pi }^2 q_2^2 \left(z \left(-3+z^2\right) q_1^3+\left(3-9 z^2\right) q_1^2 q_3-z \left(5+z^2\right) q_1 q_3^2+\left(-3-3
z^2+4 z^4\right) q_3^3\right)\right)+\nn
&&\frac{z g_A^6 M_{\pi } \left(2 M_{\pi }^2+q_3^2\right) \left(q_2^2 q_3+4 M_{\pi }^2 \left(z q_1+q_3\right)\right)}{128 F^6 \pi
 q_1 \left(-4 \left(-1+z^2\right) M_{\pi }^2+q_2^2\right) q_3^2 \left(4 M_{\pi }^2+q_3^2\right)}\,, \nn [5pt]
R_4&=&\frac{A\left(q_2\right) g_A^6 q_2^2 \left(-2 z^2
q_1^2 q_3+\left(1+z^2\right) q_3^3+2 M_{\pi }^2 \left(2 z q_1+\left(1+z^2\right) q_3\right)\right)}{128 F^6 \pi  \left(-1+z^2\right)^2 q_1^2 q_3^3}+\nn
&&\frac{A\left(q_1\right) g_A^6 \left(-2 M_{\pi }^2 \left(2 z q_1^2+\left(1+3 z^2\right) q_1 q_3+2 z q_3^2\right)+q_3 \left(2 z^2 q_1^3+2 z^3
q_1^2 q_3+\left(1-4 z^2+z^4\right) q_1 q_3^2-2 z q_3^3\right)\right)}{128 F^6 \pi  \left(-1+z^2\right)^2 q_1 q_3^3}-\nn
&&\frac{A\left(q_3\right) g_A^6}{128 F^6 \pi  \left(-1+z^2\right)^2 q_1^2 q_3^2} \left(2 M_{\pi }^2 \left(-z^2 \left(-3+z^2\right) q_1^2+z \left(3+z^2\right) q_1 q_3+\left(1+z^2\right) q_3^2\right)+\right.\nn
&&\left.q_3 \left(-\left(z+z^3\right) q_1^3-\left(1-z^2+2 z^4\right) q_1^2 q_3+z \left(1+z^2\right) q_1 q_3^2+\left(1+z^2\right) q_3^3\right)\right)-\nn
&&\frac{I(4:0,-q_1,q_3;0) g_A^6}{32 F^6 \left(-1+z^2\right)^2 q_1 \left(-4 \left(-1+z^2\right)
      M_{\pi }^2+q_2^2\right) q_3^2}\left(q_2^4 q_3 \left(\left(z+z^3\right) q_1^2+\left(-1+z^2\right)^2 q_1 q_3-2 z q_3^2\right)+\right.\nn
&&8 (-1+z) (1+z) M_{\pi }^4 \left(3 z^2 q_1^3+9 z^3 q_1^2 q_3+\left(-2+9 z^2+2 z^4\right) q_1 q_3^2+z \left(2+z^2\right) q_3^3\right)+\nn
&&\left.2 M_{\pi }^2 q_2^2 \left(z^2 \left(-3+z^2\right) q_1^3+\left(2 z-8 z^3\right) q_1^2 q_3+\left(4+5 z^2 \left(-3+z^2\right)\right)
q_1 q_3^2+2 z \left(-3+z^2+z^4\right) q_3^3\right)\right)+\nn
&&\frac{z g_A^6 M_{\pi } \left(2 M_{\pi }^2+q_3^2\right) \left(q_2^2 q_3+4 M_{\pi }^2 \left(z q_1+q_3\right)\right)}{128 F^6 \pi
 q_1 \left(-4 \left(-1+z^2\right) M_{\pi }^2+q_2^2\right) q_3^2 \left(4 M_{\pi }^2+q_3^2\right)}\,, \nn [5pt]
R_5&=&\frac{A\left(q_2\right) g_A^6 q_2^2 \left(-4 M_{\pi
}^2 \left(q_1+z q_3\right)+q_3 \left(2 z q_1^2+\left(-1+z^2\right) q_1 q_3-2 z q_3^2\right)\right)}{128 F^6 \pi  \left(-1+z^2\right)^2 q_1 q_3^4}-\nn
&&\frac{A\left(q_3\right) g_A^6 }{128 F^6 \pi  \left(-1+z^2\right)^2 q_1 q_3^3}\left(2 M_{\pi }^2 \left(z \left(-3+z^2\right) q_1^2-2 \left(1+z^2\right) q_1 q_3-2 z q_3^2\right)+q_3 \left(\left(1+z^2\right)
q_1^3+2 z^3 q_1^2 q_3-\right.\right.\nn
&&\left.\left.\left(1+z^2\right) q_1 q_3^2-2 z q_3^3\right)\right)+\nn
&&\frac{A\left(q_1\right) g_A^6 \left(2 M_{\pi }^2 \left(2 q_1^2+4 z q_1 q_3+\left(1+z^2\right) q_3^2\right)+q_3 \left(-2 z q_1^3+\left(1-3 z^2\right)
q_1^2 q_3+2 z q_1 q_3^2+\left(1+z^2\right) q_3^3\right)\right)}{128 F^6 \pi  \left(-1+z^2\right)^2 q_3^4}+\nn
&&\frac{I(4:0,-q_1,q_3;0) g_A^6}{32 F^6 \left(-1+z^2\right)^2 \left(-4 \left(-1+z^2\right) M_{\pi
      }^2+q_2^2\right) q_3^3}\left(q_2^4 q_3 \left(\left(1+z^2\right) q_1^2+z \left(-1+z^2\right) q_1 q_3-\left(1+z^2\right) q_3^2\right)+\right.\nn
&&8 (-1+z) (1+z) M_{\pi }^4 \left(3 z q_1^3+\left(-1+10 z^2\right) q_1^2 q_3+3 z \left(1+2 z^2\right) q_1 q_3^2+\left(1+2 z^2\right) q_3^3\right)+\nn
&&\left.2 M_{\pi }^2 q_2^2 \left(z \left(-3+z^2\right) q_1^3+\left(3-9 z^2\right) q_1^2 q_3-z \left(5+z^2\right) q_1 q_3^2+\left(-3-3
z^2+4 z^4\right) q_3^3\right)\right)-\nn
&&\frac{g_A^6 M_{\pi } \left(2 M_{\pi }^2+q_3^2\right) \left(q_2^2 q_3+4 M_{\pi }^2 \left(z q_1+q_3\right)\right)}{128 F^6 \pi
 \left(-4 \left(-1+z^2\right) M_{\pi }^2+q_2^2\right) q_3^3 \left(4 M_{\pi }^2+q_3^2\right)}\,, \nn [5pt]
R_6&=&\frac{A\left(q_2\right) g_A^6 \left(2 M_{\pi }^2+q_2^2\right)}{128 F^6 \pi }+\frac{A\left(q_1\right) g_A^6 \left(2 z \left(M_{\pi
}^2+q_1^2\right) q_3+q_1 \left(8 M_{\pi }^2+3 q_1^2+q_3^2\right)\right)}{128 F^6 \pi  q_1}+\nn
&&\frac{A\left(q_3\right) g_A^6 \left(2 z q_1 \left(M_{\pi }^2+q_3^2\right)+q_3 \left(8 M_{\pi }^2+q_1^2+3 q_3^2\right)\right)}{128 F^6 \pi  q_3}-\nn
&&\frac{g_A^6 M_{\pi }}{128 F^6 \pi  q_1 \left(4 M_{\pi }^2+q_1^2\right)
  \left(4 \left(-1+z^2\right) M_{\pi }^2-q_2^2\right) q_3 \left(4 M_{\pi
    }^2+q_3^2\right)}
 \left(\left(5+z^2\right) q_1^3 q_2^2 q_3^3+8 M_{\pi }^6 \left(z \left(-3+4
       z^2\right) q_1^2+\right.\right.\nn
&&\left.2 \left(19-18 z^2\right) q_1
q_3+z \left(-3+4 z^2\right) q_3^2\right)+2 M_{\pi }^4 \left(4 z
\left(-1+z^2\right) q_1^4+\left(77-36 z^2\right) q_1^3 q_3+2 z \left(33+8
  z^2\right) q_1^2 q_3^2+
\right.\nn
&&\left.\left(77-36 z^2\right)
q_1 q_3^3+4 z \left(-1+z^2\right) q_3^4\right)+\left.2 M_{\pi }^2 q_1 q_3
\left(\left(10+z^2\right) q_1^4+2 z \left(9+2 z^2\right) q_1^3 q_3+\left(29-7
    z^2\right) q_1^2 q_3^2+\right.\right.\nn
&&\left.\left.2
z \left(9+2 z^2\right) q_1 q_3^3+\left(10+z^2\right) q_3^4\right)\right)-\nn
&&\frac{I(4:0,-q_1,q_3;0) g_A^6 \left(2 M_{\pi }^2+q_2^2\right)}{32 F^6 q_1 \left(-4 \left(-1+z^2\right) M_{\pi }^2+q_2^2\right)
    q_3}\left(q_1 q_2^2 q_3 \left(q_1^2+z q_1 q_3+q_3^2\right)+4 M_{\pi }^4
    \left(z q_1^2-2 \left(-2+z^2\right) q_1 q_3+z
      q_3^2\right)+\right.\nn
&&\left.2
M_{\pi }^2 \left(4 q_1 q_3 \left(q_1^2+q_3^2\right)+z \left(q_1^4+6 q_1^2
    q_3^2+q_3^4\right)\right)\right)\,, 
\nn [5pt]
R_7&=&\frac{3 g_A^6 M_{\pi } \left(2 M_{\pi }^2+q_2^2\right)}{256 F^6 \pi  q_1^2 \left(-4 \left(-1+z^2\right) M_{\pi }^2+q_2^2\right)}-\frac{3
A\left(q_3\right) g_A^6 \left(2 M_{\pi }^2+q_2^2\right) \left(\left(1+z^2\right) q_1+2 z q_3\right)}{256 F^6 \pi  \left(-1+z^2\right)^2 q_1^3}-\nn
&&\frac{3 A\left(q_1\right) g_A^6 \left(2 M_{\pi }^2+q_2^2\right) \left(2 z q_1+\left(1+z^2\right) q_3\right)}{256 F^6 \pi  \left(-1+z^2\right)^2
q_1^2 q_3}+\frac{3 A\left(q_2\right) g_A^6 \left(2 M_{\pi }^2+q_2^2\right) \left(2 z q_1^2+\left(1+3 z^2\right) q_1 q_3+2 z q_3^2\right)}{256 F^6
\pi  \left(-1+z^2\right)^2 q_1^3 q_3}+\nn
&&\frac{3 I(4:0,-q_1,q_3;0) g_A^6 \left(2 M_{\pi }^2+q_2^2\right)}{64 F^6 \left(-1+z^2\right)^2 q_1^2 \left(4 \left(-1+z^2\right)
      M_{\pi }^2-q_2^2\right)}\left(-q_2^2 \left(\left(1+z^2\right) q_1^2+z
      \left(3+z^2\right) q_1 q_3+\left(1+z^2\right) q_3^2\right)+\right.\nn
&&\left.4 \left(-1+z^2\right) M_{\pi
}^2 \left(\left(1+2 z^2\right) q_1^2+2 z \left(2+z^2\right) q_1 q_3+\left(1+2
    z^2\right) q_3^2\right)\right)\,, 
\nn [5pt]
R_8&=&-\frac{3 z g_A^6 M_{\pi } \left(2 M_{\pi }^2+q_2^2\right)}{256 F^6 \pi  q_1 \left(-4 \left(-1+z^2\right) M_{\pi }^2+q_2^2\right)
q_3}+\frac{3 z A\left(q_3\right) g_A^6 \left(2 M_{\pi }^2+q_2^2\right) \left(\left(1+z^2\right) q_1+2 z q_3\right)}{256 F^6 \pi  \left(-1+z^2\right)^2
q_1^2 q_3}+\nn
&&\frac{3 z A\left(q_1\right) g_A^6 \left(2 M_{\pi }^2+q_2^2\right) \left(2 z q_1+\left(1+z^2\right) q_3\right)}{256 F^6 \pi  \left(-1+z^2\right)^2
q_1 q_3^2}-\frac{3 z A\left(q_2\right) g_A^6 \left(2 M_{\pi }^2+q_2^2\right) \left(2 z q_1^2+\left(1+3 z^2\right) q_1 q_3+2 z q_3^2\right)}{256 F^6
\pi  \left(-1+z^2\right)^2 q_1^2 q_3^2}-\nn
&&\frac{3 I(4:0,-q_1,q_3;0) z g_A^6 \left(2 M_{\pi }^2+q_2^2\right)}{64 F^6 \left(-1+z^2\right)^2 q_1 \left(4 \left(-1+z^2\right) M_{\pi
      }^2-q_2^2\right) q_3}\left(-q_2^2 \left(\left(1+z^2\right) q_1^2+z
      \left(3+z^2\right) q_1 q_3+\left(1+z^2\right) q_3^2\right)+\right.\nn
&&\left.4 \left(-1+z^2\right) M_{\pi
}^2 \left(\left(1+2 z^2\right) q_1^2+2 z \left(2+z^2\right) q_1 q_3+\left(1+2
    z^2\right) q_3^2\right)\right)\,, 
\nn [5pt]
R_9&=&-\frac{3 A\left(q_2\right) g_A^6 \left(2 M_{\pi }^2+q_2^2\right) \left(\left(1+z^2\right) q_1^2+z \left(3+z^2\right) q_1 q_3+\left(1+z^2\right)
q_3^2\right)}{256 F^6 \pi  \left(-1+z^2\right)^2 q_1^2 q_3^2}+\nn
&&\frac{3 A\left(q_1\right) g_A^6 \left(\left(1+z^2\right) q_1^3+2 z \left(2+z^2\right) q_1^2 q_3-z^2 \left(-7+z^2\right) q_1 q_3^2+2 z q_3^3+2
M_{\pi }^2 \left(\left(1+z^2\right) q_1+2 z q_3\right)\right)}{256 F^6 \pi  \left(-1+z^2\right)^2 q_1 q_3^2}+\nn
&&\frac{3 A\left(q_3\right) g_A^6 \left(2 z q_1^3-z^2 \left(-7+z^2\right) q_1^2 q_3+2 z \left(2+z^2\right) q_1 q_3^2+\left(1+z^2\right) q_3^3+2
M_{\pi }^2 \left(2 z q_1+\left(1+z^2\right) q_3\right)\right)}{256 F^6 \pi  \left(-1+z^2\right)^2 q_1^2 q_3}+\nn
&&\frac{3 I(4:0,-q_1,q_3;0) z g_A^6 \left(2 M_{\pi }^2+q_2^2\right)}{64 F^6 \left(-1+z^2\right)^2 q_1 \left(-4 \left(-1+z^2\right)
      M_{\pi }^2+q_2^2\right) q_3}\left(q_2^2 \left(-2 q_1^2+z
      \left(-5+z^2\right) q_1 q_3-2 q_3^2\right)+\right.\nn
&&\left.4 \left(-1+z^2\right) M_{\pi }^2 \left(\left(2+z^2\right)
q_1^2+6 z q_1 q_3+\left(2+z^2\right) q_3^2\right)\right)-\frac{3 z g_A^6 M_{\pi } \left(2 M_{\pi }^2+q_2^2\right)}{256 F^6 \pi  q_1 \left(-4 \left(-1+z^2\right) M_{\pi }^2+q_2^2\right)
q_3}\,, \nn [5pt]
R_{10}&=&\frac{3 \left(-1+z^2\right) g_A^6 M_{\pi } \left(2 M_{\pi }^2+q_2^2\right)}{256 F^6 \pi  \left(-4 \left(-1+z^2\right) M_{\pi
}^2+q_2^2\right)}+\frac{3 A\left(q_2\right) g_A^6 \left(2 M_{\pi }^2+q_2^2\right) \left(z q_1+q_3\right) \left(q_1+z q_3\right)}{256 F^6 \pi  \left(-1+z^2\right)
q_1 q_3}-\nn
&&\frac{3 A\left(q_1\right) g_A^6 \left(z q_1^3+\left(1+2 z^2\right) q_1^2 q_3-z \left(-4+z^2\right) q_1 q_3^2+q_3^3+2 M_{\pi }^2 \left(z q_1+q_3\right)\right)}{256
F^6 \pi  \left(-1+z^2\right) q_3}-\nn
&&\frac{3 A\left(q_3\right) g_A^6 \left(q_1^3-z \left(-4+z^2\right) q_1^2 q_3+\left(1+2 z^2\right) q_1 q_3^2+z q_3^3+2 M_{\pi }^2 \left(q_1+z
q_3\right)\right)}{256 F^6 \pi  \left(-1+z^2\right) q_1}+\nn
&&\frac{3 I(4:0,-q_1,q_3;0) g_A^6 \left(2 M_{\pi }^2+q_2^2\right)}{64 F^6 \left(-1+z^2\right) \left(4 \left(-1+z^2\right) M_{\pi
      }^2-q_2^2\right)}\left(-q_2^2 \left(q_1^2-z \left(-3+z^2\right) q_1
      q_3+q_3^2\right)+\right.\nn
&&\left.4 \left(-1+z^2\right) M_{\pi }^2
    \left(\left(1+z^2\right) q_1^2+4
z q_1 q_3+\left(1+z^2\right) q_3^2\right)\right)\,, \nn [5pt]
R_{11}&=&-\frac{A\left(q_2\right) g_A^6 q_2^2 \left(4 M_{\pi }^2+q_1^2+q_3^2\right)}{256 F^6 \pi  \left(-1+z^2\right) q_1^2 q_3^2}+\nn
&&\frac{A\left(q_3\right) g_A^6 \left(2 M_{\pi }^2 \left(\left(-1+z^2\right) q_1^2+2 z q_1 q_3+2 q_3^2\right)+q_3 \left(z q_1^3+\left(-1+2 z^2\right)
q_1^2 q_3+z q_1 q_3^2+q_3^3\right)\right)}{256 F^6 \pi  \left(-1+z^2\right) q_1^2 q_3^2}+\nn
&&\frac{A\left(q_1\right) g_A^6 \left(2 M_{\pi }^2 \left(2 q_1^2+2 z q_1 q_3+\left(-1+z^2\right) q_3^2\right)+q_1 \left(q_1^3+z q_1^2 q_3+\left(-1+2
z^2\right) q_1 q_3^2+z q_3^3\right)\right)}{256 F^6 \pi  \left(-1+z^2\right) q_1^2 q_3^2}-\nn
&&\frac{I(4:0,-q_1,q_3;0) g_A^6 q_2^2 }{\left(64 F^6 \left(-1+z^2\right)
q_1^2 \left(-4 \left(-1+z^2\right) M_{\pi }^2+q_2^2\right)
q_3^2\right)}\left(-\left(2 M_{\pi }^2+q_1^2\right) \left(2 M_{\pi
}^2+q_3^2\right) \left(4 M_{\pi }^2+q_1^2+q_3^2\right)+\right.\nn
&&\left.2 z^3 q_1 q_3 \left(-4 M_{\pi }^4+q_1^2
q_3^2\right)+z^2 \left(4 M_{\pi }^2+q_1^2+q_3^2\right) \left(4 M_{\pi }^4+3 q_1^2 q_3^2+2 M_{\pi }^2 \left(q_1^2+q_3^2\right)\right)+\right.\nn
&&\left.z q_1 q_3 \left(8 M_{\pi }^4+q_1^4+q_3^4+4 M_{\pi }^2
    \left(q_1^2+q_3^2\right)\right)\right)-\nn
&&\frac{g_A^6 M_{\pi } }{256 F^6 \pi  q_1^2 \left(4 M_{\pi }^2+q_1^2\right) \left(-4
      \left(-1+z^2\right) M_{\pi }^2+q_2^2\right) q_3^2 \left(4 M_{\pi
      }^2+q_3^2\right)}\left(-2 z^2
    \left(4 M_{\pi }^4 q_1^2 \left(4 M_{\pi }^2+q_1^2\right)+\right.\right.\nn
&&\left.\left.4 M_{\pi }^2
      \left(2 M_{\pi }^2+q_1^2\right)^2 q_3^2+\left.\left(2
M_{\pi }^2+q_1^2\right)^2 q_3^4\right)+2 M_{\pi }^2 \left(4 M_{\pi
}^2+q_1^2+q_3^2\right) \left(q_1^2 q_3^2+2 M_{\pi }^2
\left(q_1^2+q_3^2\right)\right)-\right.\right.\nn
&&\left.\left.z q_1 q_3 \left(32 M_{\pi }^6+12 M_{\pi }^4
\left(q_1^2+q_3^2\right)+\right.q_1^2 q_3^2 \left(q_1^2+q_3^2\right)+2 M_{\pi }^2 \left(q_1^4+4
q_1^2 q_3^2+q_3^4\right)\right)\right)\,.
\eeqa
In the above expressions, $q_1$ and $q_3$ are always to be understood as the
magnitudes of the corresponding three-momenta (except in the arguments of the
function $I$), $q_1 \equiv | \vec q_1 \, |$, 
$q_3 \equiv | \vec q_3 \, |$. Further, the function $I(d:p_1,p_2,p_3;p_4)$ 
refers to the scalar loop integral 
\beqa
I(d:p_1,p_2,p_3; p_4)&=&\frac{1}{i}\int
\frac{d^d l}{(2\pi)^d}\frac{1}{(l+p_1)^2-M_\pi^2+i \epsilon}
\frac{1}{(l+p_2)^2-M_\pi^2+i \epsilon}
\frac{1}{(l+p_3)^2-M_\pi^2+i \epsilon}\frac{1}{v\cdot(l+p_4)+i \epsilon}\,.
\quad\quad
\eeqa
In a general case, this function depends on the four-momenta $p_i$. For the
case $p_i^0 =0$ which we are interested in, it can be expressed in terms
of the three-point function in  Euclidean space 
$J\left(d:\vec{p}_1,\vec{p}_2,\vec{p}_3\right)$
\beq
J\left(d:\vec{p}_1,\vec{p}_2,\vec{p}_3\right)=
\int\frac{d^d
  l}{(2\pi)^d}\frac{1}{(\vec{l}+\vec{p}_1)^2+M_\pi^2}\frac{1}{(\vec{l}+\vec{p}_2)^2+M_\pi^2}\frac{1}{(\vec{l}+\vec{p}_3)^2+M_\pi^2}.
\eeq
In particular, the function $I\left(4:0,-q_1,q_3;0\right)$ which
enters the expressions for $R_i$ can be written as
\beq
I\left(4:0,-q_1,q_3;0\right) = \frac{1}{2}J\left(3:\vec{0},-\vec{q}_1,\vec{q}_3\right).
\eeq

For diagram (5), we obtain the following representation:
\beqa
\label{ringS}
V_{\rm ring}&=&
{\fet \tau}_1\cdot{\fet\tau}_2  \; S_1+
\vec{\sigma}_1\cdot\vec{q}_1\vec{\sigma}_3\cdot\vec{q}_1 \; {\fet\tau}_1\cdot{\fet\tau}_2
 \; S_2+
\vec{\sigma}_1\cdot\vec{q}_3\vec{\sigma}_3\cdot\vec{q}_1 \; {\fet\tau}_1\cdot{\fet\tau}_2
 \; S_3+
\vec{\sigma}_1\cdot\vec{q}_1\vec{\sigma}_3\cdot\vec{q}_3 \; {\fet\tau}_1\cdot{\fet\tau}_2
 \; S_4\nn
&+&
\vec{\sigma}_1\cdot\vec{q}_3\vec{\sigma}_3\cdot\vec{q}_3 \; {\fet\tau}_1\cdot{\fet\tau}_2
 \; S_5+
\vec{\sigma}_1\cdot\vec{\sigma}_3 \; {\fet\tau}_1\cdot{\fet\tau}_2
 \; S_6+
\vec{q}_1\cdot \vec{q}_3\times\vec{\sigma}_1  \; 
{\fet\tau}_1\cdot {\fet\tau}_2\times{\fet\tau}_3  \;  S_7 \,,
\eeqa
where the functions $S_i \equiv S_i (q_1, \, q_3,\, z)$ are given by
\beqa
S_1&=&-\frac{A\left(q_1\right) g_A^4 \left(2 M_{\pi }^2+q_1^2\right)}{128 F^6 \pi }-\frac{A\left(q_2\right) g_A^4 \left(4 M_{\pi }^2+q_1^2+z q_1
    q_3+q_3^2\right)}{128 F^6 \pi }-\frac{A\left(q_3\right)
g_A^4 \left(2 M_{\pi }^2+q_3^2\right)}{128 F^6 \pi }+\nn
&&\frac{I\left(4:0,-q_1,q_3;0\right) g_A^4 \left(2 M_{\pi }^2+q_1^2\right)
\left(2 M_{\pi }^2+q_3^2\right)}{32 F^6}-\frac{g_A^4 M_{\pi }}{64 F^6 \pi }\,,
\nn [5pt]
S_2&=&-\frac{A\left(q_1\right) g_A^4 \left(\left(1+z^2\right) q_1+2 z q_3\right)}{128 F^6 \pi  \left(-1+z^2\right)^2 q_1}-\frac{A\left(q_3\right)
g_A^4 q_3 \left(2 z q_1+\left(1+z^2\right) q_3\right)}{128 F^6 \pi  \left(-1+z^2\right)^2 q_1^2}+\nn
&&\frac{I\left(4:0,-q_1,q_3;0\right) g_A^4 q_3 \left(-4 z \left(-1+z^2\right) M_{\pi }^2+2 z q_1^2+\left(1+3 z^2\right)
q_1 q_3+2 z q_3^2\right)}{32 F^6 \left(-1+z^2\right)^2 q_1}+\nn
&&\frac{A\left(q_2\right) g_A^4 \left(\left(1+z^2\right) q_1^2+z \left(3+z^2\right) q_1 q_3+\left(1+z^2\right) q_3^2\right)}{128 F^6 \pi  \left(-1+z^2\right)^2
q_1^2}\,,\nn [5pt]
S_3&=&\frac{A\left(q_3\right) g_A^4 \left(\left(1+z^2\right) q_1+2 z q_3\right)}{128 F^6 \pi  \left(-1+z^2\right)^2 q_1}+\frac{A\left(q_1\right)
g_A^4 \left(2 z q_1+\left(1+z^2\right) q_3\right)}{128 F^6 \pi  \left(-1+z^2\right)^2 q_3}+\frac{z A\left(q_2\right) g_A^4 \left(-2 q_1^2+z \left(-5+z^2\right)
q_1 q_3-2 q_3^2\right)}{128 F^6 \pi  \left(-1+z^2\right)^2 q_1 q_3}-\nn
&&\frac{I\left(4:0,-q_1,q_3;0\right) g_A^4 \left(-4 \left(-1+z^2\right) M_{\pi }^2+\left(1+z^2\right) q_1^2+z \left(3+z^2\right)
q_1 q_3+\left(1+z^2\right) q_3^2\right)}{32 F^6 \left(-1+z^2\right)^2}\,,\nn [5pt]
S_4&=&\frac{z A\left(q_1\right) g_A^4 \left(\left(1+z^2\right) q_1+2 z q_3\right)}{128 F^6 \pi  \left(-1+z^2\right)^2 q_3}+\frac{z A\left(q_3\right)
g_A^4 \left(2 z q_1+\left(1+z^2\right) q_3\right)}{128 F^6 \pi  \left(-1+z^2\right)^2 q_1}+\nn
&&\frac{I\left(4:0,-q_1,q_3;0\right) z g_A^4 \left(4 z \left(-1+z^2\right) M_{\pi }^2-2 z q_1^2-\left(1+3 z^2\right)
q_1 q_3-2 z q_3^2\right)}{32 F^6 \left(-1+z^2\right)^2}-\nn
&&\frac{z A\left(q_2\right) g_A^4 \left(\left(1+z^2\right) q_1^2+z \left(3+z^2\right) q_1 q_3+\left(1+z^2\right) q_3^2\right)}{128 F^6 \pi  \left(-1+z^2\right)^2
q_1 q_3}\,,\nn [5pt]
S_5&=&-\frac{A\left(q_1\right) g_A^4 q_1 \left(\left(1+z^2\right) q_1+2 z q_3\right)}{128 F^6 \pi  \left(-1+z^2\right)^2 q_3^2}-\frac{A\left(q_3\right)
g_A^4 \left(2 z q_1+\left(1+z^2\right) q_3\right)}{128 F^6 \pi  \left(-1+z^2\right)^2 q_3}+\nn
&&\frac{I\left(4:0,-q_1,q_3;0\right) g_A^4 q_1 \left(-4 z \left(-1+z^2\right) M_{\pi }^2+2 z q_1^2+\left(1+3 z^2\right)
q_1 q_3+2 z q_3^2\right)}{32 F^6 \left(-1+z^2\right)^2 q_3}+\nn
&&\frac{A\left(q_2\right) g_A^4 \left(\left(1+z^2\right) q_1^2+z \left(3+z^2\right) q_1 q_3+\left(1+z^2\right) q_3^2\right)}{128 F^6 \pi  \left(-1+z^2\right)^2
q_3^2}\,,\nn [5pt]
S_6&=&-\frac{A\left(q_3\right) g_A^4 q_3 \left(z q_1+q_3\right)}{128 F^6 \pi  \left(-1+z^2\right)}+\frac{A\left(q_2\right) g_A^4 \left(q_1^2-z \left(-3+z^2\right) q_1
    q_3+q_3^2\right)}{128 F^6 \pi  \left(-1+z^2\right)}-\frac{A\left(q_1\right) g_A^4 q_1
\left(q_1+z q_3\right)}{128 F^6 \pi  \left(-1+z^2\right)}+\nn
&&\frac{I\left(4:0,-q_1,q_3;0\right) g_A^4 q_1 q_3 \left(z q_1+q_3\right)
\left(q_1+z q_3\right)}{32 F^6 \left(-1+z^2\right)}\,,\nn [5pt]
S_7&=&\frac{A\left(q_1\right) g_A^4 \left(2 M_{\pi }^2+q_3^2\right)}{256 F^6 \pi  \left(-1+z^2\right) q_3^2}-\frac{A\left(q_2\right) g_A^4 \left(z q_3^2 \left(z q_1+q_3\right)+2 M_{\pi }^2 \left(q_1+z q_3\right)\right)}{256 F^6 \pi  \left(-1+z^2\right)
q_1 q_3^2}+\frac{z A\left(q_3\right)
g_A^4 \left(2 M_{\pi }^2+q_3^2\right)}{256 F^6 \pi  \left(-1+z^2\right) q_1
q_3}-\nn
&&\frac{I\left(4:0,-q_1,q_3;0\right) g_A^4
\left(z q_1+q_3\right) \left(2 M_{\pi }^2+q_3^2\right)}{64 F^6 \left(-1+z^2\right) q_3}\,.
\eeqa

Examining the above results one observes that the individual terms in the
expressions for $R_i$ and $S_i$ are singular for $z=\pm 1$, $q_1=0$ and/or $q_3=0$. 
These singularities, however, cancel in such a way that the
resulting terms in Eqs.~(\ref{ringR}) and (\ref{ringS}) are finite. 
In principle, it is possible to obtain a representation for functions $R_i$
and $S_i$ which is free of at least some of the singularities. In particular, the
singularities at $z=\pm 1$ can be avoided if one expresses the results 
in terms of the functions $J_1$ and $J_2$ defined as 
\beqa
J_1(d,\vec{q}_1,\vec{q}_3)&=&\frac{1}{1-z^2}\Big\{J(d:\vec{0},-\vec{q}_1,\vec{q}_3)-\frac{1}{2}(1+z)
\Big[\frac{J(d:\vec{0},\vec{q}_1)}{q_3^2+q_1 q_3}+\frac{J(d:\vec{0},\vec{q}_3)}{q_1^2+q_1 q_3}
-\frac{J(d:\vec{0},\vec{q}_1+\vec{q}_3)}{q_1 q_3}\Big]\nonumber\\
&-&\frac{1}{2}(1-z)\Big[\frac{J(d:\vec{0},\vec{q}_1)}{q_3^2-q_1 q_3}+
\frac{J(d:\vec{0},\vec{q}_3)}{q_1^2-q_1 q_3}+\frac{J(d:\vec{0},\vec{q}_1+\vec{q}_3)}{q_1 q_3}\Big]\Big\},\\
J_2(d,\vec{q}_1,\vec{q}_3)&=&\frac{1}{(1-z^2)^2}\Big\{J(d:\vec{0},-\vec{q}_1,\vec{q}_3)
-\frac{1}{4}(1-z)^2\Big[\frac{J(d:\vec{0},\vec{q}_1)}{q_3^2-q_1 q_3}
+\frac{J(d:\vec{0},\vec{q}_3)}{q_1^2-q_1 q_3}+\frac{J(d:\vec{0},\vec{q}_1+\vec{q}_3)}{q_1 q_3}\\
&+&(1+z)\Big[-\frac{8 M^2 -2 q_1^2+(d-1)(q_1 - q_3)(2 q_1 - q_3)}{(d-1)q_3(q_1-q_3)^3}J(d:\vec{0},\vec{q}_1)\nonumber\\
&+&\frac{8 M^2-2 q_3^2 + (d-1)(q_1-q_3)(q_1-2 q_3)}{(d-1)q_1(q_1-q_3)^3}
J(d:\vec{0},\vec{q}_3)
+\frac{2(4 M^2 +(d-2)(q_1-q_3)^2)}{(d-1)q_1 q_3
  (q_1-q_3)^2}J(d:\vec{0},\vec{q}_1+\vec{q}_3)\Big]
\Big]\nonumber\\
&-&\frac{1}{4}(1+z)^2\Big[\frac{J(d:\vec{0},\vec{q}_1)}{q_3^2+q_1 q_3}
+\frac{J(d:\vec{0},\vec{q}_3)}{q_1^2+q_1 q_3}-\frac{J(d:\vec{0},\vec{q}_1+\vec{q}_3)}{q_1 q_3}\nonumber\\
&+&(1-z)\Big[\frac{8 M^2 - 2 q_1^2 + (d - 1)(q_1+q_3)(2 q_1 + q_3)}
{(d-1)q_3(q_1+q_3)^3}J(d:\vec{0},\vec{q}_1)\nonumber\\
&+&
\frac{8 M^2 - 2 q_3^2 + (d-1)(q_1 + q_3)(q_1 + 2 q_3)}{(d-1)q_1(q_1+q_3)^3}
J(d:\vec{0},\vec{q}_3)
-\frac{2(4 M^2 +(d-2)(q_1 + q_3)^2)}{(d-1)q_1 q_3 (q_1+q_3)^2}J(d:\vec{0},\vec{q}_1+\vec{q}_3)
\Big]\Big]\Big\},\nonumber
\eeqa
rather than the three-point function $J(d:\vec{0},-\vec{q}_1,\vec{q}_3 )$ and
uses certain linear combinations of two-point functions
and tadpole integrals. In the above expressions, the two-point function is
defined as 
\beq
J\left(d:\vec{p}_1,\vec{p}_2\right)=
\int\frac{d^d l}{(2\pi)^d}\frac{1}{(\vec{l}+\vec{p}_1)^2+M_\pi^2}\frac{1}{(\vec{l}+\vec{p}_2)^2+M_\pi^2}\,. 
\eeq

%


\end{document}